\documentclass[a4paper,fleqn,usenatbib]{mnras}
\usepackage{amsmath}
\usepackage{amssymb}
\usepackage{cite}
\usepackage{natbib}
\usepackage{graphicx} 
\usepackage{xcolor}
\usepackage{txfonts}

\usepackage[T1]{fontenc}
\usepackage{ae,aecompl}

\usepackage{etoolbox}
\makeatletter
\newcount\c@additionalboxlevel
\newcount\c@maxboxlevel
\patchcmd\@combinedblfloats{\box\@outputbox}{%
  \stepcounter{additionalboxlevel}%
  \box\@outputbox
}{}{\errmessage{\noexpand\@combinedblfloats could not be patched}}

\AtBeginShipout{%
  \ifnum\value{additionalboxlevel}>\value{maxboxlevel}%
    \typeout{Warning: maxboxlevel might be too small, increase to %
      \the\value{additionalboxlevel}%
    }%
  \fi 
  \@whilenum\value{additionalboxlevel}<\value{maxboxlevel}\do{%
    \typeout{* Additional boxing of page `\thepage'}%
    \setbox\AtBeginShipoutBox=\hbox{\copy\AtBeginShipoutBox}%
    \stepcounter{additionalboxlevel}%
  }%
 \setcounter{additionalboxlevel}{0}%
}
\makeatother

\renewcommand\vec{\bmath}

\title{%
AMICO galaxy clusters in KiDS-DR3: weak-lensing mass calibration
}

\author[Bellagamba et al.]{%
Fabio Bellagamba,$^{1,2}$ Mauro Sereno,$^{1,2}$ Mauro Roncarelli,$^{1,2}$ Matteo Maturi,$^{4}$  
\newauthor Mario Radovich,$^{3}$ Sandro Bardelli,$^{2}$ Emanuella Puddu,$^{5}$ Lauro Moscardini,$^{1,2,6}$
\newauthor Fedor Getman,$^{5}$ Hendrik Hildebrandt$^{7}$ and Nicola Napolitano$^{5,8}$\\
$^1$Dipartimento di Fisica e Astronomia, Alma Mater Studiorum Universit\`a di Bologna, via Gobetti 93/2, I-40129 Bologna, Italy\\
$^2$INAF - Osservatorio di Astrofisica e Scienza dello Spazio di Bologna, via Gobetti 93/3, I-40129 Bologna, Italy\\
$^3$INAF - Osservatorio Astronomico di Padova, vicolo dell'Osservatorio 5, Padova 35122, Italy\\
$^4$Zentrum f\"ur Astronomie, Universit\"at Heidelberg, Philosophenweg 12, D-69120 Heidelberg, Germany\\
$^5$INAF - Osservatorio Astronomico di Capodimonte, Salita Moiariello 16, Napoli 80131, Italy\\
$^6$INFN - Sezione di Bologna, Viale Berti Pichat 6/2, I-40127 Bologna, Italy\\
$^7$Argelander-Institut f\"ur Astronomie, Auf dem H\"ugel 71, 53121 Bonn, Germany\\
$^8$School of Physics and Astronomy,  Sun Yat-sen University Zhuhai campus,  2 Daxue Road,  Tangjia,  Zhuhai,  Guangdong 519082,  P.R. China
}

\pubyear{2018}

\begin{document}
\label{firstpage}
\pagerange{\pageref{firstpage}--\pageref{lastpage}}
\maketitle

\begin{abstract}
We present the mass calibration for galaxy clusters detected with the AMICO code in KiDS DR3 data. The cluster sample comprises $\sim$ 7000 objects and covers the redshift range 0.1 < $z$ < 0.6. We perform a weak lensing stacked analysis by binning the clusters according to redshift and two different mass proxies provided by AMICO, namely the amplitude $A$ (measure of galaxy abundance through an optimal filter) and the richness $\lambda^\text{\textasteriskcentered}$ (sum of membership probabilities in a consistent radial and magnitude range across redshift). For each bin, we model the data as a truncated NFW profile plus a 2-halo term, taking into account uncertainties related to concentration and miscentring. From the retrieved estimates of the mean halo masses, we construct the $A$-$M_{200}$ and the $\lambda^\text{\textasteriskcentered}$-$M_{200}$ relations. The relations extend over more than one order of magnitude in mass, down to $M_{200} \sim 2\, (5) \times 10^{13} M_\odot/h$ at $z$ = 0.2\,(0.5), with small evolution in redshift. The logarithmic slope is $\sim 2.0$ for the $A$-mass relation, and $\sim 1.7$ for the $\lambda^\text{\textasteriskcentered}$-mass relation, consistent with previous estimations on mock catalogues and coherent with the different nature of the two observables.
\end{abstract}

\begin{section}{Introduction}
Ongoing and upcoming photometric surveys will increase the current census of clusters of galaxies by orders of magnitude, pushing the limits of detection towards lower masses and higher redshifts. In order to exploit this amount of information for astrophysics and cosmology, it is fundamental to know how to relate an appropriate measure of galaxies, accessible in photometric observations, to the total mass of clusters, mostly made by dark matter. Unfortunately, the relation between the total mass and the galaxy properties in a cluster is hidden in complex astrophysical processes, the details of which are difficult to model theoretically or through simulations. For this reason, there is not an obvious mass proxy in the galaxy distribution and the scaling between the mass and any galaxy observable can only be calibrated empirically. Many different mass proxies have been suggested in the literature, mainly based on the number of (red) galaxies inside a given radius \citep{2010MNRAS.404.1922A,2012ApJ...746..178R}, their luminosities \citep{2014MNRAS.443.3309M} or photometric stellar mass estimates \citep{2018MNRAS.474.1361P}.
 
Galaxy clusters are detected in photometric data by running algorithms which search for significant cluster-scale overdensities in the galaxy distribution. In this procedure, usually one or more observables related to the abundance and properties of galaxies are measured and used in the identification of the clusters. Although it is possible to measure a posteriori any mass proxy from the observed galaxy distribution, it is natural to study the mass-observable relation for these quantities that are automatically included in the output cluster catalogue, as has been done for example by \citet{2015MNRAS.452..701W}, \citet{2017ApJ...848..114P} and \citet{2017MNRAS.466.3103S}. Moreover, the selection function of the sample with respect to the observable(s) used in its definition will likely be simpler to assess and is usually one of the fundamental results of the procedure.

To obtain reliable mass estimates for the clusters, it is convenient to take advantage of weak gravitational lensing, the imprint on background galaxies' observed shapes of the light deflection due to the intervening cluster potential. Gravitational lensing depends on the total matter distribution in the cluster and provides mass estimates which do not rely on any assumption on the physical state of the clusters, differently from methods which depend on the gas properties, such as X-ray observations or the Sunyaev-Zel'dovich effect on the CMB. Moreover, gravitational lensing is often one of the primary objectives of current and future photometric surveys, thus allowing the calibration of optical observables of photometrically-selected clusters within the same data-set. 

In this paper we focus on clusters detected in the Kilo-Degree Survey \citep[KiDS,][]{2013Msngr.154...44J} by AMICO (Adaptive Matched Identifier of Clustered Objects), an optimal filtering algorithm presented in \citet{2018MNRAS.473.5221B} and recently selected as an official detection algorithm for the Euclid survey \citep[][Adam et al. in prep.]{2011arXiv1110.3193L}. The Kilo Degree Survey is an ESO public survey being performed with the OmegaCAM wide-field camera \citep{2011Msngr.146....8K} mounted at the VLT Survey Telescope \citep[VST,][]{2011Msngr.146....2C}. KiDS is designed to observe a total area of 1500 sq. degrees in the \textit{ugri} bands. The third Data Release (DR3) includes 440 sq. degrees \citep{2017A&A...604A.134D}. In \citet{2018arXiv181002811M}, a catalog of $\sim$ 8000 clusters at $0 < z < 0.8$ detected by AMICO in KiDS DR3 data will be presented and validated. 

The primary scope of this paper is to determine the calibration between the main observable measured by AMICO, namely the amplitude, and the cluster masses. The amplitude is a measure of the cluster galaxy content in units of the input model, which is constructed from known data. In \citet{2018MNRAS.473.5221B} it has been shown that the amplitude is a reliable mass proxy on simulations, provided the model calibration is correct. In the detection procedure, AMICO also calculates the probability for each galaxy to be a member of each detected structure. We will take advantage of this feature to build a second optical observable for the same catalog, as the sum of membership probabilities in a given radial and magnitude range, and we will also investigate its properties as a mass proxy.

Weak lensing analyses of KiDS data have already been performed to constrain cosmological parameters through cosmic shear \citep{2017MNRAS.465.1454H}, to derive density profiles of GAMA galaxy groups \citep{2015MNRAS.452.3529V} and of their satellites \citep{2015MNRAS.454.3938S}, the stellar-to-halo mass relation in galaxies \citep{2016MNRAS.459.3251V} and to study throughs and ridges in the galaxy density field \citep{2018arXiv180500562B}.  In this work, we will use KiDS shear data to build excess surface density profiles for clusters detected by AMICO in the same photometric data. As the weak lensing signal-to-noise ratio for each object is typically very small, we will use a stacking approach and measure the mean profile and mass of ensembles of clusters selected according to their optical properties. We will then derive the mass-observable relations for both our optical mass proxies, taking into account the systematic uncertainties in the selection of background galaxies, in the photometric redshifts estimate and in the shear measurement.

The paper is organised as follows. In Section 2 we present the two data sets used in this analysis: the cluster catalogue and the galaxy catalogue. In Section 3 we detail the method used to extract the stacked differential density profiles of the lenses from galaxy shear estimates. The matter distribution used to model the lenses, and the parameters on which it depends, are described in Section 4. In Section 5 we outline how the parameters are derived from the comparison between data and model. In Section 6 we quantify the systematic uncertainties that affect our analysis, and in Section 7 we describe the method to derive parameters of the mass-observable relation from the weak lensing results. In Section 8 we show the results for the weak lensing profile of each cluster bin, and for the mass-observable relations in our cluster sample. Some considerations on the different observables employed and their relation with previous works in literature are given in Section 9. A summary and our final remarks are presented in Section 10.

For the purpose of this analysis, we assume a flat $\Lambda$CDM cosmology with $\text{H}_0$ = 70 Km/s/Mpc and $\Omega_\text{m}$ = 0.3. Halo masses are given as $M_{200,\text{c}}$, the mass enclosed in a sphere with radius $r_{200,\text{c}}$, where the mean density is 200 times the critical density of the Universe at the corresponding redshift.
\end{section}

\begin{section}{Data}
Our work is based on data from KiDS Data Release 3 \citep{2017A&A...604A.134D}, covering 440 tiles with an area of $\sim$1 sq. degree each, and observed in all four survey filters ($ugri$). KiDS consists of two patches, one in the equatorial sky (KiDS-N) and the other around the South Galactic Pole (KiDS-S), and DR3 includes complete coverage of the Northern GAMA fields \citep{2011MNRAS.413..971D}. The limiting $5\sigma$-magnitudes in the four bands are respectively 24.20, 25.09, 24.96 and 23.62, in a 2 arcsec aperture. The data processing and catalog extraction are done by a KiDS-optimised pipeline running in the Astro-WISE environment \citep{2011arXiv1112.0886V,2013ExA....35...45M}. 

\begin{figure}
 \includegraphics[width=\columnwidth]{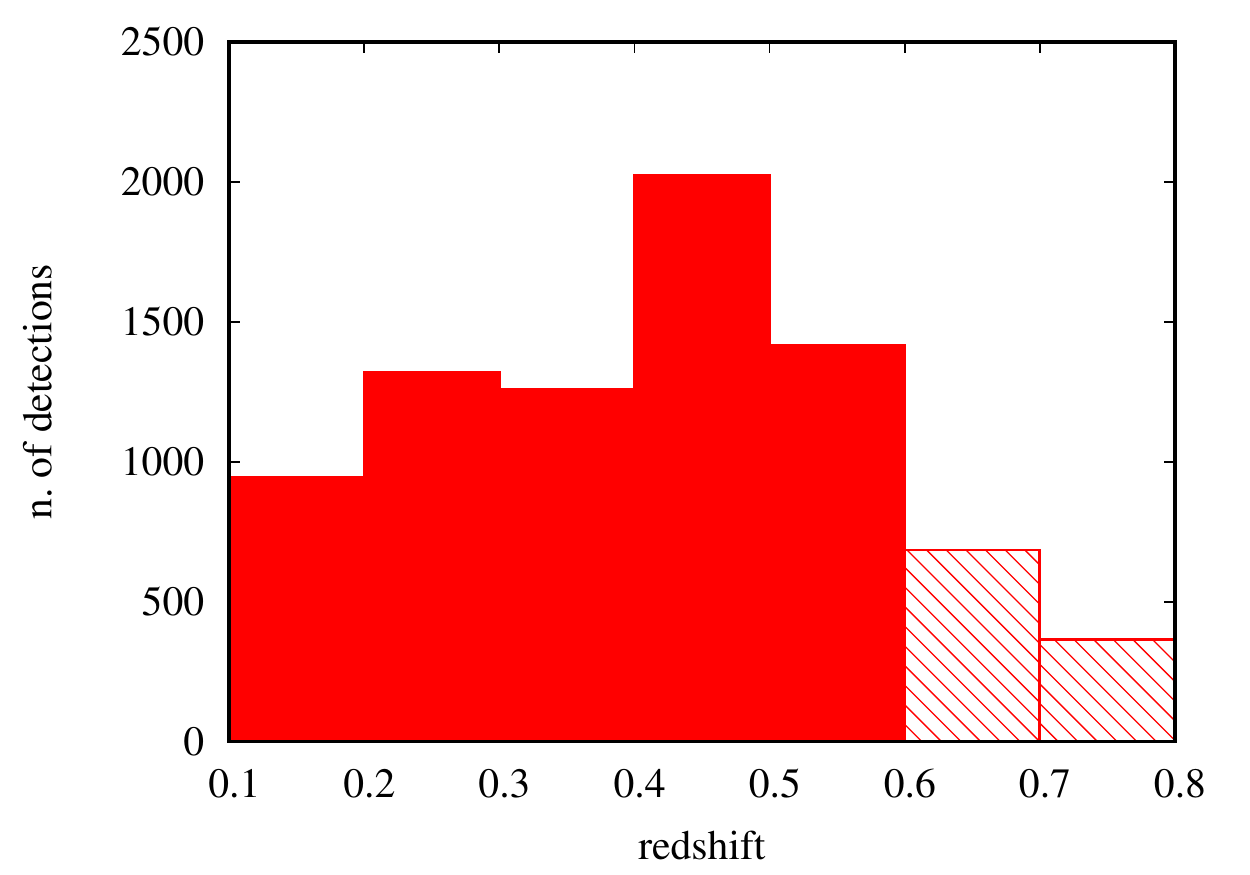}
 \caption{Redshift distribution of the cluster sample extracted by AMICO on KiDS DR3 data. The striped bins indicate clusters not used in this analysis.}
 \label{fig:cluster_z_distr}
\end{figure}

\begin{figure}
 \includegraphics[width=\columnwidth]{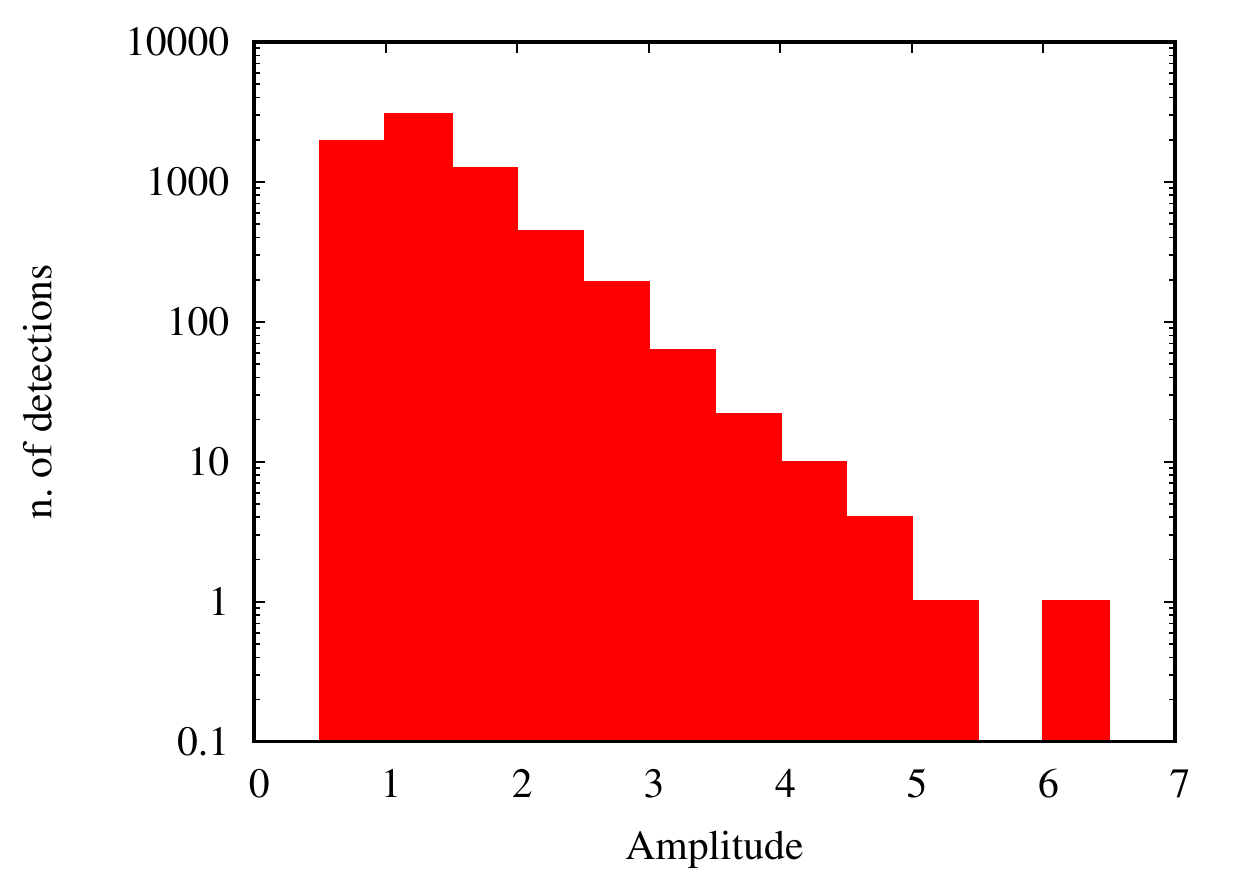}
 \caption{Ampltude distribution of the cluster sample used in the analysis, extracted by AMICO on KiDS DR3 data. As the selection is performed with a S/N threshold , the amplitude threshold is not constant in redshift. This produces the increase in cluster counts from the first to the second bin in amplitude.}
 \label{fig:cluster_amp_distr}
\end{figure}

\begin{subsection}{Cluster catalog}\label{sect:clu_cat}
The catalog of galaxy clusters has been extracted from KiDS DR3 data with AMICO, an optimal filtering algorithm which has been presented and tested on simulations in \citet{2018MNRAS.473.5221B}. A full description and validation of the cluster catalogue can be found in \citet{2018arXiv181002811M}. The cluster detection with AMICO on DR3 is an improvement of the work by \citet{2017A&A...598A.107R} on DR2, both in terms of total area (440 against 114 sq. degrees) and for the evolution in the detection algorithm with respect to the previous matched filter method \citep{2011MNRAS.413.1145B}. AMICO looks for cluster candidates by convolving the 3D galaxy distribution with a redshift-dependent filter, constructed as the ratio of a cluster and a noise model. This convolution generates a 3D amplitude map, whose peaks represent the detections. For each detection, AMICO provides the angular position, redshift, S/N and amplitude. The amplitude is a measure of cluster galaxy abundance in units of the cluster model, defined as
\begin{equation}\label{eq:amplitude}
A(\vec \theta_\mathrm{c}, z_\mathrm{c})  \equiv \alpha^{-1}(z_\mathrm{c})\ \sum_{i=1}^{N_{gal}} \frac {M_\mathrm{c}(\vec \theta_i - \vec \theta_\mathrm{c}, m_i)\ p_i(z_\mathrm{c})}{N(m_i,z_\mathrm{c})} - B(z_\mathrm{c}) ,
\end{equation}
where $M_\text{c}$ is the cluster model (expected density of galaxies per unit magnitude and solid angle) at redshift $z_c$, $N$ is the noise distribution,  $p_i(z)$ , $\vec \theta_i$ and $m_i$ are the photometric redshift distribution, the sky coordinates and the magnitude of the $i$-th galaxy, respectively, and $\alpha$ and $B$ are redshift-dependent functions that provide the correct normalisation and subtraction of the background. The cluster model $M_\text{c}$ is constructed by a luminosity function and a radial profile, following the formalism shown in \citet{2018MNRAS.473.5221B}. Its parameters have been extracted from the observed galaxy population of SZ-detected clusters \citep{2017MNRAS.467.4015H}, as detailed in \citet{2018arXiv181002811M}. Moreover, AMICO assigns to each galaxy a probability to be a member of a given detection according to
\begin{equation}\label{eq:member_prob_corr}
P( i\in j ) \equiv P_{\mathrm{f},i}  \times \frac {A_j\, M_j(\vec \theta_i - \vec \theta_j, m_i)\, p_i(z_j)}{A_j\, M_j(\vec \theta_i - \vec \theta_j, m_i)\, p_i(z_j) + N(m_i,z_j)} ,
\end{equation}
where $A_j$, $\vec \theta_j$ and $z_j$ are the amplitude, the sky coordinates and the redshift of the $j$-th detection, respectively, and $P_{f,i}$ is the field probability of the $i$-th galaxy before the $j$-th detection is defined (see \citet{2018MNRAS.473.5221B} for all details).

In the application to KiDS data, AMICO was run considering galaxy coordinates, $r$-band magnitude and the full photometric $p(z)$ distribution. The complete catalog comprises 8092 candidate clusters at redshifts $z$ < 0.8. In the following, we will consider only clusters in the redshift range 0.1 $\leq z$ < 0.6. We discarded objects at $z$ < 0.1 because of the reduced lensing power and those at $z \ge$ 0.6 because the density of background galaxies in KiDS data does not allow a robust lensing analysis. This selection leaves us with 6962 objects. Their distribution in redshift and amplitude is shown in Figures \ref{fig:cluster_z_distr} and \ref{fig:cluster_amp_distr}. In the following, we will divide the cluster sample in three redshift bins: 0.1 $\leq z$ < 0.3, 0.3 $\leq z$ < 0.45, 0.45 $\leq z$ < 0.6. We applied to the cluster redshifts a correction of $-0.02(1+z)$, to remove the bias found in \citet{2018arXiv181002811M}.
\end{subsection}

\begin{subsection}{Galaxy catalog}\label{sect:gal_cat}
Sources that are located behind clusters appear distorted by the gravitational potential of the intervening matter. We are therefore interested in their shapes (which encode the information about the lens matter distribution) and their redshifts (which weigh the lensing information). Here we will review the main properties of the catalogue. For a more thorough discussion of the shear and redshift properties of the galaxy sample we refer the reader to \citet{2017A&A...604A.134D} and \citet{2017MNRAS.465.1454H}.

\begin{subsubsection}{Shape measurements}
The shear analysis of KiDS data was presented and discussed in \citet{2015MNRAS.454.3500K} and \citet{2017MNRAS.465.1454H}. The shape measurement was done with \textsc{lensfit} \citep{2007MNRAS.382..315M,2013MNRAS.429.2858M,2017MNRAS.467.1627F}, a likelihood based model-fitting method which has been successfully used in the analyses of other datasets, such as CFHTLenS \citep{2013MNRAS.429.2858M} and the Red Cluster Survey \citep{2016MNRAS.463..635H}. Only \textit{r}-band data have been used for shear measurement, as they are the ones with better seeing properties and highest source density. The multiplicative shear calibration error estimated from simulations is on the order of 1 per cent \citep{2017MNRAS.467.1627F}. The final catalogue provides shear measurements for $\sim$ 15 million galaxies, with an effective number density of $n_{\text{eff}}$ = 8.53 galaxies $\text{arcmin}^{-2}$ \citep[following the definition by][]{2012MNRAS.427..146H} over a total effective area of 360.3 $\deg^2$.
\end{subsubsection}

\begin{subsubsection}{Photometric redshifts}
The properties of KiDS photometric redshifts have been presented and discussed in \citet{2015MNRAS.454.3500K} and \citet{2017A&A...604A.134D}. Photometric redshifts from 4-band (\textit{ugri}) data have been derived with the Bayesian template-fitting method \textsc{BPZ} \citep{2000ApJ...536..571B,2012MNRAS.421.2355H}. When compared with spectroscopic redshifts from GAMA \citep{2015MNRAS.452.2087L}, the resulting accuracy is $\sigma_z \sim 0.04 (1+z)$, as shown in \citet{2017A&A...604A.134D}. 
\end{subsubsection}
\end{subsection}

\end{section}

\begin{section}{Weak lensing profile}\label{sect:profile}
\begin{subsection}{Selection of background sources}\label{sect:sel_back}
In order to extract the shear profile of a given lens, we first need to select the background galaxies, i.e. sources that lie behind the lens and whose observed shape is perturbed by the intervening mass distribution. If galaxies belonging to the cluster or in the cluster foreground are mistakenly considered as background, the measured lensing signal can be significantly diluted \citep[see e.g.][]{2005ApJ...619L.143B,2007ApJ...663..717M}. A possible approach is to consider all galaxies as suitable background sources, weigh them according to their redshift probability distribution, and then account separately for the inevitable inclusion of some cluster members and foreground sources \citep{2014MNRAS.442.1507G,2017MNRAS.469.4899M}. We instead choose to follow a more conservative approach described in \citet{2017MNRAS.472.1946S}, which aims at excluding foreground and cluster galaxies from the catalog, performing cuts based on photometric redshift distributions and colours. To this aim, a first selection is performed excluding the galaxies whose most likely redshift is not significantly higher than the lens one. Galaxies with photometric redshift $z_\text{s}$ pass this cut if
\begin{equation}
z_{\text{s}} > z_{\text{l}} + \Delta z ,
\end{equation}
where $z_l$ is the redshift of the lens and $\Delta z$ is set to 0.05, similar to the typical uncertainty on photometric redshifts in the galaxy catalog and well larger than the uncertainty on the cluster redshifts.

A galaxy enters the background catalogue if it passes this preliminary selection and fulfils at least one of the two following criteria, based on photo-$z$s and on colours respectively.
\begin{subsubsection}{Photo-z selection}
The criterion based on the photometric redshift distribution aims at keeping only galaxies which have a well-behaved $p(z)$ and have a negligible probability of being at redshift equal to or lower than the cluster. So the following selections are made:
\begin{itemize}
\item $z_{\text{s,min}} > z_{\text{l}} + \Delta z$
\item $\mathrm{ODDS} \ge 0.8$
\item $0.2 \leq z_{\text{s}} \leq 1.0$
\end{itemize}
The value of $z_{\text{s,min}}$ is the lower bound of the region including the $2\sigma$ (95.4 per cent) of the probability density distribution. By requiring that it is bigger than $z_\text{l} + \Delta z$ we exclude galaxies whose nominal redshift is higher than the lens but have a non-negliglble probability of being at a redshift lower than or equal to the one of the cluster. The ODDS parameter quantifies which fraction of the $p(z)$ is concentrated around the peak value, making this threshold useful in excluding distributions with significant secondary solutions or wide tails \citep{2006AJ....132..926C,2012MNRAS.422..553B}. The third selection encloses the range for which we expect the most reliable photometric redshift estimations given the available set of bands ($ugri$) and thus excludes galaxies whose redshift estimation is less robust. 

\end{subsubsection}
\begin{subsubsection}{Colour selection}
The drawback of the photo-z selection described above is the exclusion of a significant fraction of galaxies located behind the cluster, that would be useful to increase the shear signal, due to the stringent requirements on the properties of the $p(z)$ distribution. We recover part of this population, while keeping a very low fraction of contaminants, by invoking a selection based on galaxy colours, as done for example by \citet{2010MNRAS.405..257M} and \citet{2017arXiv170600434M}. Specifically, the cut we perform is the following:
\begin{displaymath}
(g - r < 0.3) \ \vee \ ( r - i > 1.3 ) \ \vee \ (r-i > g-r).
\end{displaymath}
This cut was originally proposed by \citet{2012MNRAS.420.3213O} based on the properties of the galaxies in the COSMOS catalogue \citep{2009ApJ...690.1236I} to select galaxies at $z$ $\gtrsim$ 0.7, and has been subsequently tested and used by \citet{2014ApJ...784L..25C},\citet{2017MNRAS.472.1946S,2018NatAs...2..744S}. In particular, \citet{2017MNRAS.472.1946S} showed that the 97\% of the galaxies with high quality spectroscopic redshifts selected by this cut in the CFHTLS-W1 and W4 fields have $z_{\mathrm{spec}}$ > 0.63.
\end{subsubsection}

\begin{figure}
 \includegraphics[width=\columnwidth]{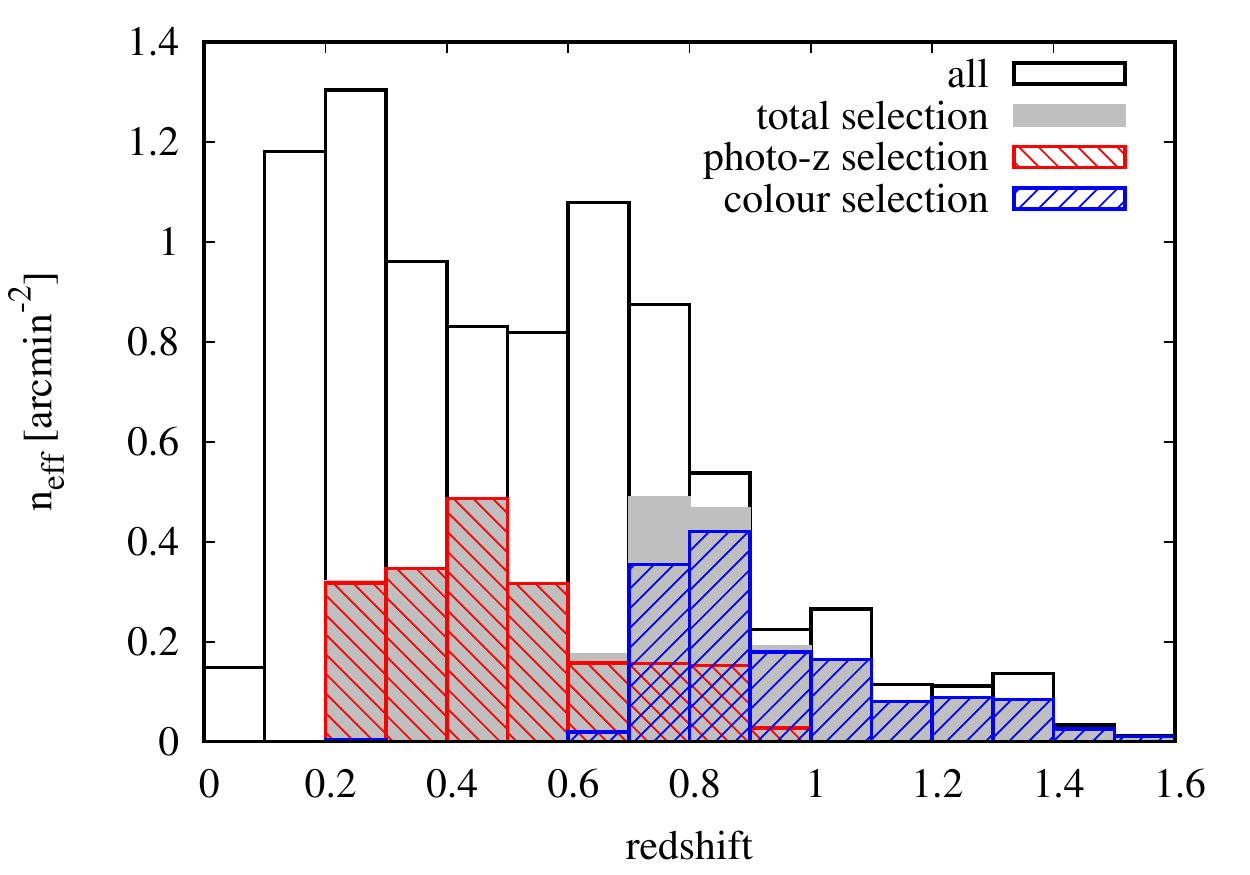}
 \caption{Effective number density as a function of $z_{\text{s}}$ of the total shear catalog (black) and of the selected background sources (filled grey). The red and blue histograms indicate the selected samples according to the two criteria employed, based on photo-$z$s and on colours, respectively. This plot considers the source most probable redshift only. For a more realistic view on the source redshift distribution, see Section \ref{sect:syst_back_pz}.}
 \label{fig:backsel_hist}
\end{figure}

This selection leaves us with an effective number density $\sim 3.15$ arcmin$^{-2}$ of available background galaxies. The distribution as a function of $z_\mathrm{s}$ of the selected sources can be seen in Figure \ref{fig:backsel_hist}. As expected, the two selection criteria are mostly complementary: most of the photo-$z$ selected sources lie at $z$ < 0.6, while the sources selected by colours have a significant tail at $z$ > 1.
\end{subsection}

\begin{subsection}{Measuring the shear profiles}\label{sect:shear_prof}

The shear is linked to the differential surface density profile of the lens via \citep{2004AJ....127.2544S}
\begin{equation}\label{eq:delta_sigma}
\Delta \Sigma (R) = \bar \Sigma (< R) - \Sigma (R) = \Sigma_{\text{crit}} \gamma_+ ,
\end{equation}
where $\gamma_+$ is the tangential component of the shear, $\Sigma (R)$ is the mass surface density, $\bar \Sigma (< R)$ is its mean inside radius $R$,  and $\Sigma_{\text{crit}}$ is the critical density for lensing \citep{2001PhR...340..291B},
\begin{equation}\label{eq:sigma_crit}
\Sigma_{\text{crit}} \equiv \frac {c^2} {4 \pi G} \frac {D_\text{s}}{D_\text{l} D_\text{ls}},
\end{equation}
that depends on the angular diameter distances between observer and source ($D_\text{s}$), observer and lens ($D_\text{l}$), and lens and source ($D_\text{ls}$).

In practice, for each lens we consider all the background galaxies selected as in Section \ref{sect:sel_back}, and we compute the shear signal in the tangential direction with respect to the center of the cluster. We stress that the shear $\gamma$ is actually unaccessible to observations. What we measure instead is an estimate of the reduced shear $g = \gamma/(1-\kappa)$, where $\kappa \equiv \Sigma/\Sigma_\text{crit}$ \citep{2001PhR...340..291B}. Anyway, in the so-called \textit{weak lensing regime} where $\kappa \ll 1$, we can approximate $\gamma \sim g$.
This allows us to construct the observed $\Delta \Sigma$ profile by computing
\begin{equation}\label{eq:delta_sigma_sum}
\Delta \Sigma (R_j) = \bigg( \frac {\sum_{i \in j} \ (w_i \Sigma_{\text{crit},i}^{-2}) \ \gamma_{+,i} \ \Sigma_{\text{crit},i}} {\sum_{i \in j} \ (w_i \Sigma_{\text{crit},i}^{-2})} \bigg) \ \frac 1 {1 + K(R_j)},
\end{equation}
where $j$ is the considered radial annulus with mean radius $R_j$, and $w_i$ is the weight assigned to the measurement of the source ellipticity \citep{2004AJ....127.2544S}. The function $K(R_j)$ is the average correction due to the multiplicative noise bias in the shear estimate and is computed from
\begin{equation}
K(R_j) = \frac {\sum_{i \in j} \ (w_i \Sigma_{\text{crit},i}^{-2}) \ m_i} {\sum_{i \in j} \ (w_i \Sigma_{\text{crit},i}^{-2})} \,
\end{equation}
where $m_i$ is the multiplicative noise bias for the $i$-th galaxy (See \citet{2017MNRAS.467.1627F} for details.)

To compute the critical density for the $i$-th galaxy $\Sigma_{\text{crit},i}$, we used the most probable redshift of the source given by BPZ. It is possible to use the full $p(z)$, and marginalise over the real source redshift to derive an effective (inverse) $\Sigma_{\text{crit}}$ as done, for example, by \citet{2004AJ....127.2544S} or \citet{2015MNRAS.452.3529V}. However, this procedure assumes an exact knowledge of the full $p(z)$, including the tails of the distribution, which are often poorly determined, as shown by \citet{2017MNRAS.465.1454H}. We will anyway use the full $p(z)$ to estimate the systematic uncertainty related to the source redshift in Section \ref{sect:syst_unc}. We will instead ignore the uncertainty in the lens redshift, which is typically $\sim 0.02(1+z)$ \citep{2018arXiv181002811M} and whose impact on $\Sigma_{\text{crit}}$ is negligible with respect to the source one.

In the analysis, the excess surface density was computed for 14 logarithmically spaced radial bins in physical units from 0.1 to 3.16 $\text{Mpc}/h$. 
\end{subsection}

\begin{subsection}{Stacking}\label{sect:stacking}
For the vast majority of the clusters in the sample, the signal-to-noise ratio in the lensing data is too low to constrain the density profile, and thus to measure a reliable mass. For this reason, we will measure the mean mass for ensembles of objects selected according to their observables and their redshift, as will be detailed in Section \ref{sect:results}. To this end, we define a weighted sum of the lensing signal for classes of objects as follows.

The differential density profile for the $K$-th cluster bin is estimated as
\begin{equation}
\Delta \Sigma_{K}(R_j) = \frac {\sum_{k \in K} W_{k,j} \ \Delta \Sigma_{k}(R_j)} {\sum_{k \in K} W_{k,j}},
\end{equation}
where $k$ runs over the clusters in the bin, and $W_{k,j}$ is the total weight for the $j$-th radial bin of the $k$-th cluster,
\begin{equation}
W_{k,j} = \sum_i w_i \Sigma_{\text{crit},i}^{-2}
\end{equation}
and $i$ runs over the background galaxies in the $j$-th radial bin. 

The covariance matrix for the stacked signal is estimated with a bootstrap procedure with replacement, as we expect the error to be dominated by shape noise. We resampled the source catalog 10000 times.
\end{subsection}

\end{section}

\begin{section}{Lens model}\label{sect:model}

\begin{figure}
 \includegraphics[width=\columnwidth]{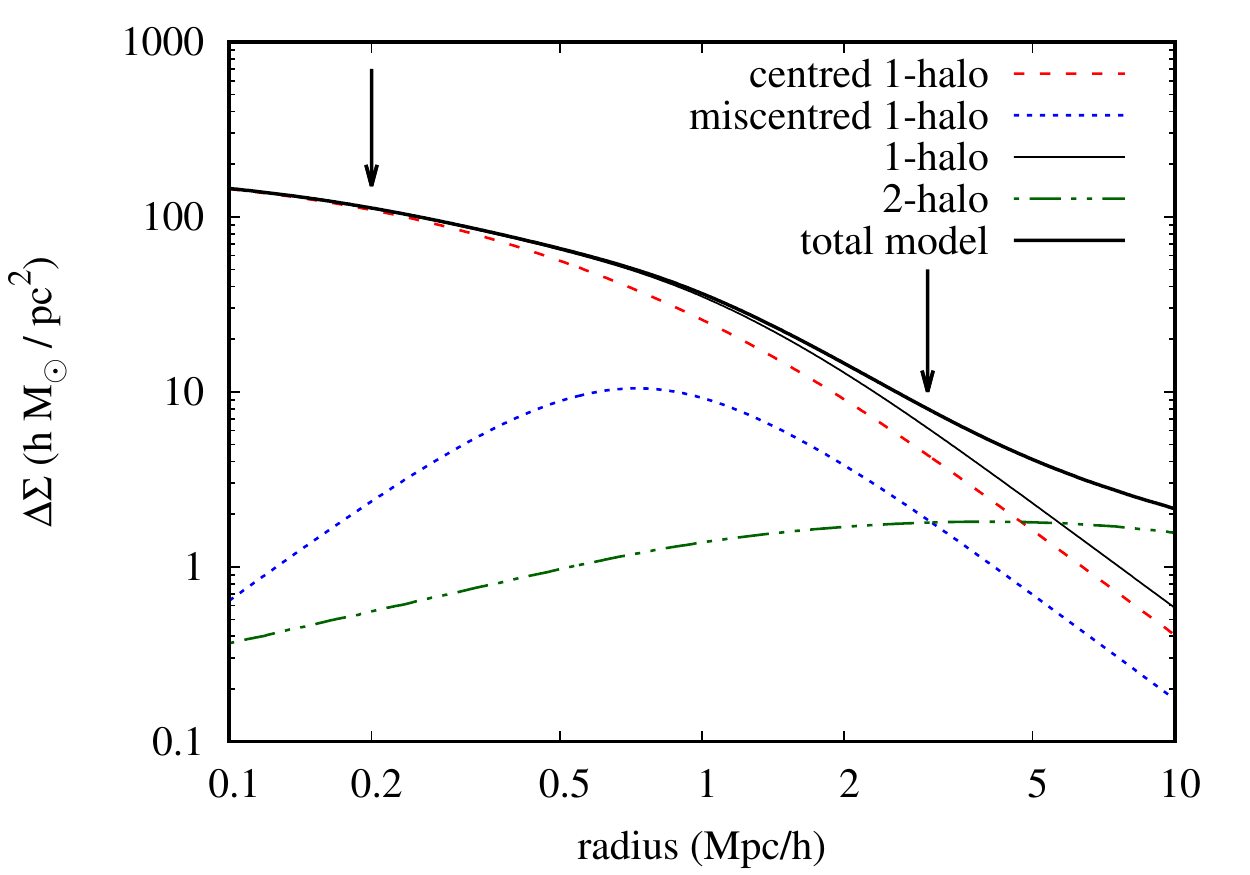}
 \caption{Differential projected density profile for a model with parameters $M_{200}$ = $10^{14} M_\odot/h$, $c$ = 3., $f_{\text{off}}$ = 0.3, $\sigma_{\text{off}}$ = 0.3 \text{Mpc}/h, at redshift $z$ = 0.3. The arrows indicate the minimum and maximum radius used for the comparison with data.}
 \label{fig:deltasigma_model}
\end{figure}

We model the stacked mass density profile of our clusters with a smoothly-truncated Navarro Frenk White (NFW) distribution \citep{2009JCAP...01..015B}
\begin{equation}\label{eq:bmo}
\rho(r) = \frac {\rho_{\text{s}}} {r/r_{\text{s}} \ (1 + r/r_{\text{s}})^2} \ \bigg( \frac{r_{\text{t}}^2} {r^2 + r_{\text{t}}^2} \bigg)^2,
\end{equation}
where $\rho_{\text{s}}$ is the typical density, $r_{\text{s}}$ is the scale radius and $r_{\text{t}}$ is the truncation radius. The scale radius is usually parametrised as $r_{\text{s}} = r_{200}/c_{200}$, where $c_{200}$ is the concentration parameter. The normalisation of the above distribution can be expressed in terms of $M_{200}$. The truncated version of the NFW is found to better describe cluster profiles outside the viral radius in simulations than the original NFW profile \citep{2011MNRAS.414.1851O}. For our analysis, we set $r_{\text{t}} = 3 r_{200}$, following \citet{2017MNRAS.472.1946S}.

In addition to the halo profile described by Equation \ref{eq:bmo}, we expect a contribution to the density profile due to matter in correlated halos. This is the so-called \textit{2-halo} term and can be modelled as \citep{2011PhRvD..83b3008O,2017MNRAS.472.1946S}
\begin{equation}\label{eq:2-halo}
\Sigma_{\text{2h}}(\theta; M,z) = \int \frac {l dl} {2 \pi} J_0(l\theta) \frac {\bar \rho_{\text{m}} (z)\, b_{\text{h}}(M;z)} {(1+z)^3\, D^2_{\text{l}} (z)} P_{\text{lin}}(k_{\text{l}};z),
\end{equation}
where $\theta$ is the angular radius, $J_0$ is the zero-th order Bessel function and $k_{\text{l}} = l/(1+z)/D_{\text{l}}(z)$. This component depends on the halo bias $b_{\text{h}}$, for which we follow the prescription of \citet{2010ApJ...724..878T} and on the linear power spectrum $P_{\text{lin}}$, which we computed according to \citet{1999ApJ...511....5E}.

A typical source of bias in cluster weak lensing analyses is the misidentification of the center of the lens. In our analysis, we consider the position of the detection made by AMICO as the center of the cluster. As the detection is performed on a grid, there is an intrinsic uncertainty related to the AMICO pixel size, which is < 0.1 Mpc/$h$ \citep{2018MNRAS.473.5221B}. Moreover, the center of the galaxy distribution may show some significant offset with respect to the center of mass \citep[see e.g.][]{2012ApJ...757....2G}. A better knowledge of the miscentring distribution of AMICO clusters would help

We parametrise this uncertainty by considering a model component produced by halos whose position has a non-negligible scatter with respect to the assumed centre, as done for example in \citet{2007arXiv0709.1159J} and in \citet{2015MNRAS.452.3529V}. First of all, we define $\sigma_{\text{off}}$ as the rms of the distribution of the misplacement of the halos on the plane of the sky. Assuming a Gaussian distribution, the probability of a lens being at distance $R_s$ from the assumed centre is then
\begin{equation}\label{eq:p_rs}
P(R_{\text{s}}) = \frac {R_{\text{s}}} {\sigma_{\text{off}}^2} \exp \bigg[-\frac 1 2 \bigg( \frac {R_{\text{s}}} {\sigma_{\text{off}}} \bigg)^2 \bigg].
\end{equation}
The azimuthally averaged profile of a population of halos misplaced by a distance $R_s$ is given by \citep{2006MNRAS.373.1159Y}
\begin{equation}\label{eq:sigma_rs}
\Sigma (R | R_{\text{s}}) = \frac 1 {2 \pi} \int_0^{2 \pi} \Sigma_{\text{cen}} \, \bigg( \sqrt {R^2 + R_{\text{s}}^2 + 2 R R_{\text{s}} \cos \theta} \bigg) \, d \theta,
\end{equation}
where $\Sigma_{\text{cen}}(R)$ is the centred surface brightness distribution derived by Equations \ref{eq:bmo} and \ref{eq:2-halo}. Finally, the mean surface density distribution derived from a mis-centred population of halos is obtained by integrating Eq. \ref{eq:sigma_rs} along $R_{\text{s}}$ and weighing each offset distance according to Eq. \ref{eq:p_rs}:
\begin{equation}
\Sigma_{\text{off}}(R) = \int P(R_{\text{s}})\, \Sigma (R | R_{\text{s}})\, d R_{\text{s}} .
\end{equation}

Our model for the halo can then be written as the sum of a centred population and an off-centred one:
\begin{equation}\label{eq:sigma_1h}
\Sigma_{\text{1h}}(R) = (1 - f_{\text{off}}) \Sigma_{\text{cen}} (R) + f_{\text{off}} \Sigma_{\text{off}} (R),
\end{equation}
where $f_{\text{off}}$ is the fraction of halos that belong to the miscentred population. The subdivision of the model in a perfectly centred component and a miscentred one aims at capturing the uncertainty linked to the centering of the halos. Other choices are possible, such as considering the AMICO pixel scale as an intrinsic smoothing scale. Future studies about the miscentring distribution of AMICO clusters would help to improve this aspect of the model.

This model depends on four parameters ($M_{200}, c_{200}, \sigma_{\text{off}}, f_{\text{off}}$). An example of the total profile can be seen in Figure \ref{fig:deltasigma_model}. As we limit our analysis to the central $\sim$ 3 $\text{Mpc}/h$, the 2-halo term affects to a small degree only the last radial bins. This makes our results less dependent on the details of the halo bias model. 

Even though we measure the shear signal from a radial distance of 0.1 $\text{Mpc}/h$, in the analysis we will neglect the first three radial bins, which is equivalent to consider only the range 0.2 $\lesssim R \lesssim$ 3 $\text{Mpc}/h$. The reason for this is twofold. First of all, from the observational point of view, measuring shear and photo-$z$ close to the cluster centre is made difficult by the cluster galaxies contamination. Furthermore, the analysis of the shear signal close to the cluster center is sensitive to the contribution of the BCG to the matter distribution and to deviations from the weak-lensing approximation used in the above derivation. This choice also alleviates the effects of miscentring.

\end{section}

\begin{section}{Inference of model parameters}\label{sect:der_par}
Given a set of data $\vec D$ of the stacked differential surface density profile $\Delta \Sigma_{\text{obs}} (R)$ derived as in Section \ref{sect:profile} and a model $\Delta \Sigma_{\text{mod}} (R)$ defined as in Section \ref{sect:model}, the posterior probability of a set of parameters $\vec \theta =  (M_{200},c_{200},f_{\text{off}},\sigma_{\text{off}})$ is given by Bayes' theorem
\begin{equation}
P(\vec \theta | \vec D ) = \frac {\mathcal{L} (\vec D | \vec \theta) P (\vec \theta)}{E(\vec D)},
\end{equation}
where $\mathcal{L} (\vec D | \vec \theta)$ is the likelihood of the data given the model, $P (\vec \theta)$ is the prior probability of the parameters and $E(\vec D)$ is the so-called \textit{evidence}, that does not depend on the parameters and that we can ignore in this context. The likelihood in our problem is given by
\begin{equation}\label{eq:like}
\mathcal{L} \propto \exp \bigg(- \frac 1 2\, \chi^2\bigg) \ ,
\end{equation}
where 
\begin{equation}
\chi^2 = \sum_{i=1}^{N_{\text{bin}}} \sum_{j=1}^{N_{\text{bin}}} [\Delta \Sigma_{\text{obs}} (R_i) - \Delta \Sigma_{\text{mod}} (R_i) ]\, \vec C^{-1}_{ij}\, [\Delta \Sigma_{\text{obs}} (R_j) - \Delta \Sigma_{\text{mod}} (R_j) ] ,
\end{equation}
$C$ is the covariance matrix and $N_\text{bin}$ is the number of radial bins in the considered profile. The main source of off-diagonal terms in the covariance matrix is correlated shape noise due to the fact that the same galaxies enter different radial bins in different cluster profiles \citep{2013MNRAS.432.1544M,2015MNRAS.449.4147S,2015MNRAS.452.3529V}. We expect this term to be fairly small in our case, as the mean radial distance between cluster centres is larger than the limiting radius for shear profiles ($\sim 3 \text{Mpc}/h$). In fact, the off-diagonal terms of the resulting $C_{ij}$ are generally compatible with random noise over a null signal. We therefore prefer to consider only the diagonal terms in our analysis, to avoid complications related to the inversion of a noisy estimate of the covariance matrix \citep[see][]{2007A&A...464..399H}. The values of the diagonal terms $C_{ii}$ scale with radius as $R^{-2}$ as expected for shape noise variance in logarithmically spaced bins.

In order to derive the posterior distribution of parameters from a given stacked profile, we use \textsc{MultiNest} \citep{2009MNRAS.398.1601F,2013arXiv1306.2144F}, a code that implements a Nested Sampling algorithm. This method allows an efficient exploration of the posterior without the need for an initial guess or a burn-in phase as in other Monte Carlo techniques. At first, a set of points is drawn from the prior distribution, then at each iteration one point from the sample is substituted by one with higher probability. The procedure stops when the uncertainty in the evidence estimation (and so, in the knowledge of the posterior) is below a user-defined threshold. 

In our analysis we assumed the following priors for the parameters:
\begin{itemize}
\item $\log (M_{200} / (M_\odot/h))$ uniform between 12.5 and 15.5
\item $c_{200}$ uniform between 1 and 20
\item $f_{\text{off}}$ uniform between 0 and 0.5
\item $\sigma_{\text{off}}$ uniform between 0 and 0.5 $\text{Mpc}/h$
\end{itemize}
These priors are meant to be conservative. For what concerns miscentring, in \citet{2018MNRAS.473.5221B} the typical miscentring of AMICO detections on simulated data was < 0.1 Mpc/$h$. Studies that compare optical detections to the center of X-ray emission typically find $f_{\text{off}} \sim 0.2 - 0.3$ and $\sigma_{\text{off}} \sim$ 0.2  Mpc/$h$ \citep{2016ApJS..224....1R,2018PASJ...70S..20O}.

The main result of a \textsc{MultiNest} run is a sample of the $n$-dimensional posterior distribution, from which different estimators and statistics can be computed. In the following, for each parameter, we quote the mean and the standard deviation of the marginalised posterior distribution as the resulting typical value and its uncertainty.

Given the intrinsic degeneracy between the concentration $c_{200}$ and the miscentring described by $\sigma_{\text{off}}$ and $f_{\text{off}}$, these three parameters remain substantially unconstrained by the data. Therefore, we will concentrate on the results about $M_{200}$ in the following. The complex modelling described in Section \ref{sect:model} is anyway necessary to parametrise our ignorance about the other parameters and marginalise over their possible values. In this way we minimise the risk of biasing our results about $M_{200}$, and we obtain realistic errors that take into account uncertainties related to modelling and miscentring.
\end{section}

\begin{section}{Systematic uncertainties}\label{sect:syst_unc}

Apart from the statistical uncertainty derived from the posterior distribution of $M_{200}$, there are other sources of uncertainty that should be considered when deriving the mass-observable relation from weak lensing data. In particular, three aspects of the analysis are critical when measuring the signal: the selection of background galaxies, the photo-$z$ estimates and the shear measurement. Uncertainties in the results of these procedures affect directly the measurement of $\Delta \Sigma$ from the data, so in order to derive their impact on the mass estimation we should consider the relation between the lensing mass and the excess density profile. Following \citet{2017MNRAS.469.4899M}, one can define the logarithmic dependence of $\Delta \Sigma$ on $M_{200}$ as
\begin{equation}\label{eq:gamma}
\Gamma = \frac {d \log \Delta \Sigma (M_{200})}{d \log M_{200}}\,.
\end{equation}
Then we can relate the uncertainties on the mass to those on $\Delta \Sigma$ as
\begin{equation}\label{eq:gamma}
\frac {\delta M_{200}} {M_{200}} \sim \frac 1 \Gamma \frac {\delta \Delta \Sigma}{\Delta \Sigma}.
\end{equation}
In the range of radii and redshifts considered by our analysis, assuming a realistic concentration $c$ = 4, we verified that $\Gamma \sim 0.75$ is a good approximation.

Other systematic uncertainties are related to the modelling of the signal, i.e. the way a measured $\Delta \Sigma$ is translated into $M_{200}$. In the following we discuss the choice of the analytic model, the non-spherical symmetry of the stacked sample and the projections by aligned halos.

\begin{subsection}{Background selection and photo-$z$s}\label{sect:syst_back_pz}
\begin{subsubsection}{Preliminary estimates}\label{sect:syst_prel}
If any galaxy at a redshift equal to or lower than the cluster enters its shear profile, it lowers the estimate of the signal, because its shape is uncorrelated with the matter distribution in the cluster. This is strictly true only if there are no intrinsic alignments, see \citet{2003MNRAS.339..711H}. Due to our conservative selection of background galaxies, we do not expect a significant contamination from foreground galaxies. For what concerns the photo-$z$ cut, by definition, given the constraint $z_{\text{s,min}} > z_\text{l}$ we expect that only 2.3\% of the galaxies lie at a redshift lower or equal to that of the cluster. Assuming idealised Gaussian distributions, the additional $\Delta z$ (= 0.05) buffer, of the order of $\sigma_z$, means that $\sim 3 \sigma$ of the p(z) are above $z_\text{l}$, and reduces the fraction of expected contaminants to $\sim 0.15 \%$. In case of biased and/or non-Gaussian distributions, the resulting contamination can be higher. 

As per the colour-colour selection, \citet{2017MNRAS.472.1946S} found that in the CFHTLS galaxy sample only 3\% of the selected galaxies with spectroscopic redshifts was at $z < 0.63$, which is higher than the upper limit for our cluster sample (0.6). We can thus conservatively consider the foreground contamination to be $\lesssim$ 2\%, which according to Equation \ref{eq:gamma} translates to a 2.7\% uncertainty on mass. 

When computing the critical density for each cluster-source pair (Equation \ref{eq:sigma_crit}), we use the most likely redshift for the source $z_s$, neglecting the uncertainty related to the photometric redshift measurement. We prefer not to deal with the low-probability tails of the $p(z)$, which are often not accurate and may influence severely the retrieved $\Sigma_{\text{crit}}$ for some galaxies. Nevertheless, the typical width and shape of the photometric redshift distributions can be used to estimate the uncertainty in the excess density profile due to photo-$z$ estimates. We do this by defining the integrated inverse critical density as
\begin{equation}\label{eq:mean_sigma_crit}
\langle \Sigma_{\text{crit}}^{-1} \rangle = \int dz_s\, p(z_s)\, \Sigma_{\text{crit}}^{-1} (z_s, z_l) .
\end{equation}
We then compute for the selected sample of background galaxies two estimates of $\Sigma_{\text{crit}}$: one derived with a point estimate of $z_\text{s}$ from Equation \ref{eq:sigma_crit} and one obtained by using the full $p(z)$ and inverting Equation \ref{eq:mean_sigma_crit}. For each cluster redshift bin, we can compare the two values of the mean $\Sigma_{\text{crit}}$ for the background sample, considering the distribution of cluster redshifts $z_\text{l}$ inside the bin. The relative difference between the two estimates for each of the three cluster redshift bins is respectively 0.041, 0.055 and 0.029. We use the mean of these three values (4.2\%) as an estimate of the systematic uncertainty related to the redshift distribution of the background sources. This estimate considers effects both from bias and scatter in the photo-$z$ peaks.

In order to further assess our estimates on the systematics related to background selection and photo-$z$s, we use two external catalogues: the composite spectroscopic catalogue used in \citet{2017MNRAS.465.1454H} to calibrate the redshift distribution of the KiDS shear sample and the COSMOS 30-band photo-$z$ catalogue \citep{2016ApJS..224...24L}.
\end{subsubsection}

\begin{subsubsection}{Comparison with spec-$z$ catalogue}\label{sect:comp_spec}
We follow \citet{2017MNRAS.465.1454H} who calibrated the redshift distribution of the KiDS shear sample with a spectroscopic sub-sample, using the method presented in \citet{2008MNRAS.390..118L}. We use the same technique to check the assumed redshift distribution of our selected background against the spectroscopic one. As the spectroscopic coverage is not uniform in magnitude space, each spectroscopic object is assigned a weight which depends on the magnitude-space distributions of the photometric sample and of the spectroscopic sub-sample. Spec-$z$ objects are up-weighted in regions of magnitude space where they are under-represented with respect to the full photometric sample and down-weighted in regions of magnitude space where they are over-represented. In this way, the effect of the selection function of spectroscopic objects is suppressed.

\begin{figure*}
\includegraphics[width=\columnwidth]{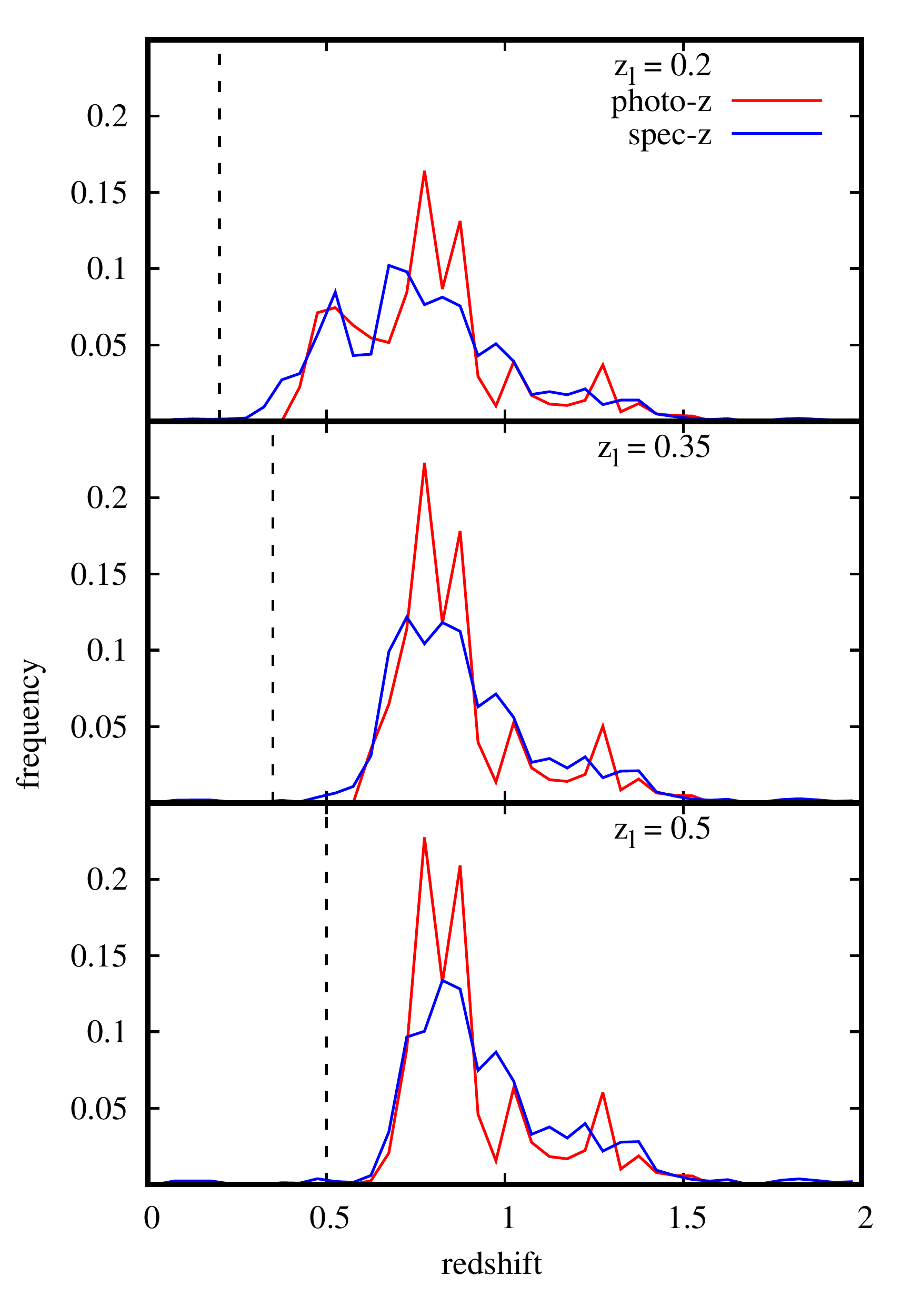}
 \includegraphics[width=\columnwidth]{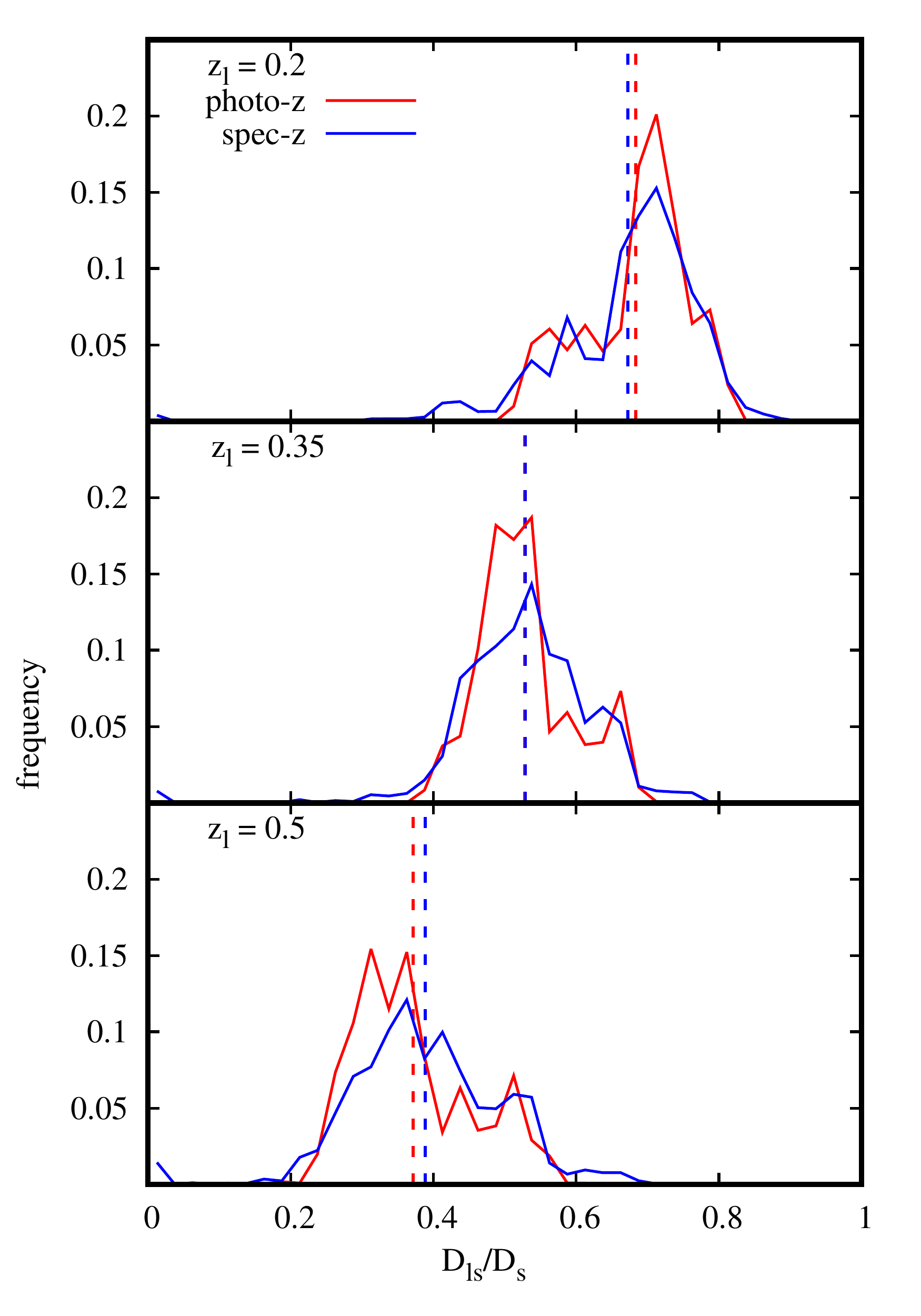}
 \caption{Left: redshift distributions of the background sample, as estimated from photometric redshifts (red line) or weighted spectroscopic redshifts of a sub-sample of objects (blue line). The black dashed line represents the assumed cluster redshift, which is 0.2, 0.35 and 0.5 for the top, middle and bottom panel, respectively. Right: distribution of $D_\text{ls}/D_\text{s}$ for the background sample, as estimated from photometric redshifts (red line) or weighted spectroscopic redshifts of a sub-sample of objects (blue line). The dashed lines represent the mean $D_\text{ls}/D_\text{s}$ for the sample. The top, middle and bottom panels refer to different cluster redshifts, with $z_\text{l}$ = 0.2, 0.35 and 0.5, respectively.}
 \label{fig:comp_spec}
\end{figure*}

As our background selection depends on the lens redshift $z_\text{l}$, we perform the comparison for three cluster redshifts, typical for the redshift bins used in the analysis: $z_l$ = 0.2, 0.35 and 0.5. For each redshift, we perform in the spectroscopic sub-sample the same selection we did on the full shear sample, according to $p(z)$ and colours. We can then compare the weighted redshift distribution of the spectroscopic sub-sample with the photo-$z$ distribution of the shear sample. Note that both distributions account for the \textsc{lensfit} weights $w$ introduced in Section \ref{sect:shear_prof}: they are used explicitly in the photo-$z$ distribution and they enter the calculation of the magnitude-space weights for the spec-$z$ objects. The results are shown in the left panel of Figure \ref{fig:comp_spec}. We can see that for all three values of $z_l$ there is a broad overlap between the two distributions, with the spec-$z$ one which is slightly less peaked at $0.7 < z < 0.9$ and has a larger tail at $z$ > 0.9. 

Following \citet{2018PASJ...70...30M}, we can estimate the contamination by foreground galaxies from the fraction of spec-$z$ distribution which lies at $z$ < $z_\text{l}$. We obtain a contamination fraction of 0.4\%, 0.7\% and 1.6\% for the three lens redshifts, respectively.

The source redshift enters the lensing analysis via the critical density $\Sigma_\text{crit}$ (see Equation \ref{eq:sigma_crit}), which, for fixed lens redshift, depends on the ratio of angular diameter distances $D_\text{ls}/D_\text{s}$. We then translate the previous redshift distributions to distributions of $D_\text{ls}/D_\text{s}$. They are shown in the right panel of Figure \ref{fig:comp_spec}, again for the same three lens redshifts $z_l$ = 0.2, 0.35 and 0.5. The mean value of $D_\text{ls}/D_\text{s}$ for the two samples differ by 1.6\%, 0.2\% and 4.3\%, respectively. This result is consistent with the systematic uncertainty due to the source redshift distributions we derived in Section \ref{sect:syst_prel} (4.2\%) from photometric data only. 

\end{subsubsection}

\begin{subsubsection} {Comparison with COSMOS photo-$z$ catalogue}\label{sect:comp_cosmos}

\begin{figure*}
\includegraphics[width=\columnwidth]{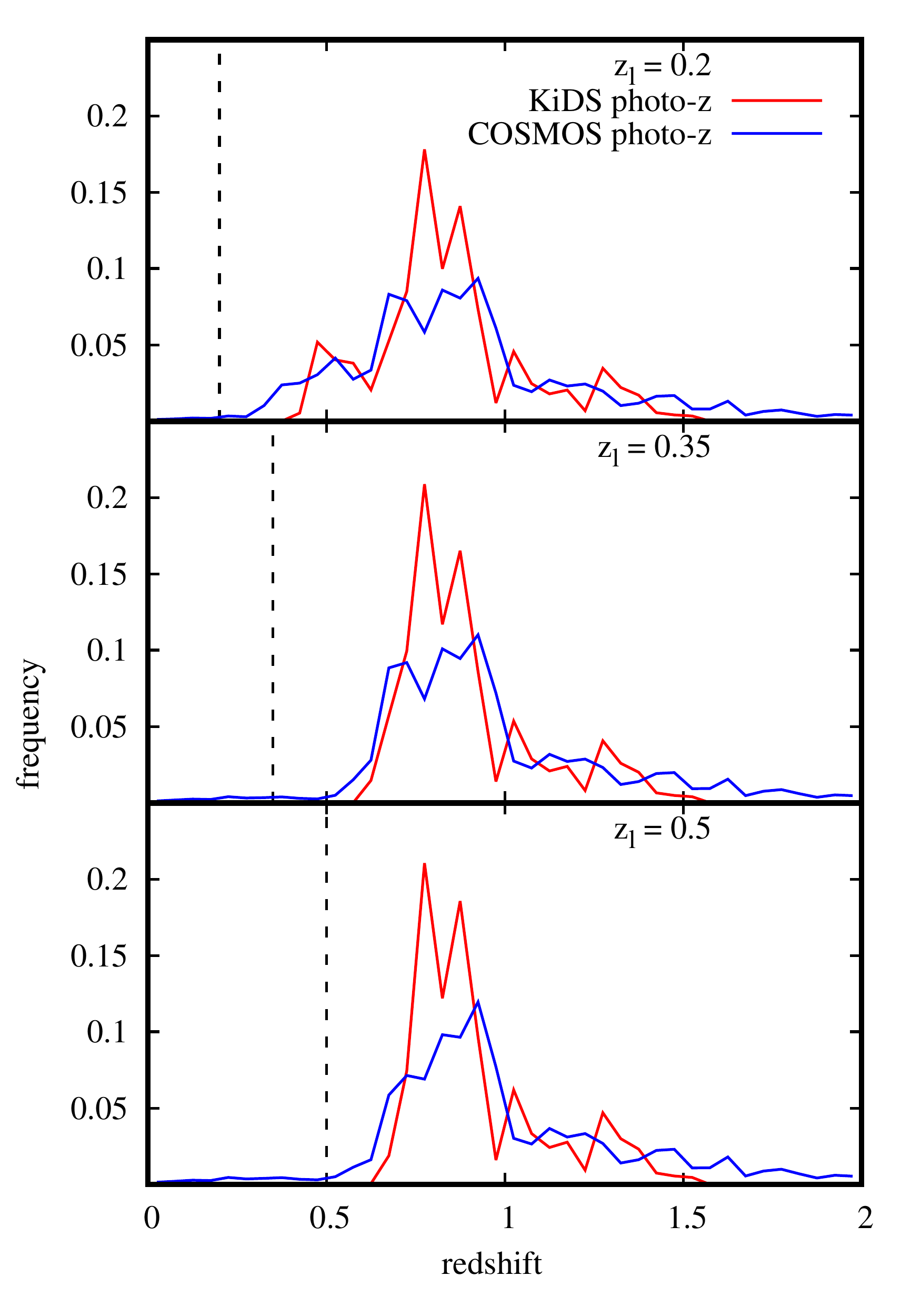}
  \includegraphics[width=\columnwidth]{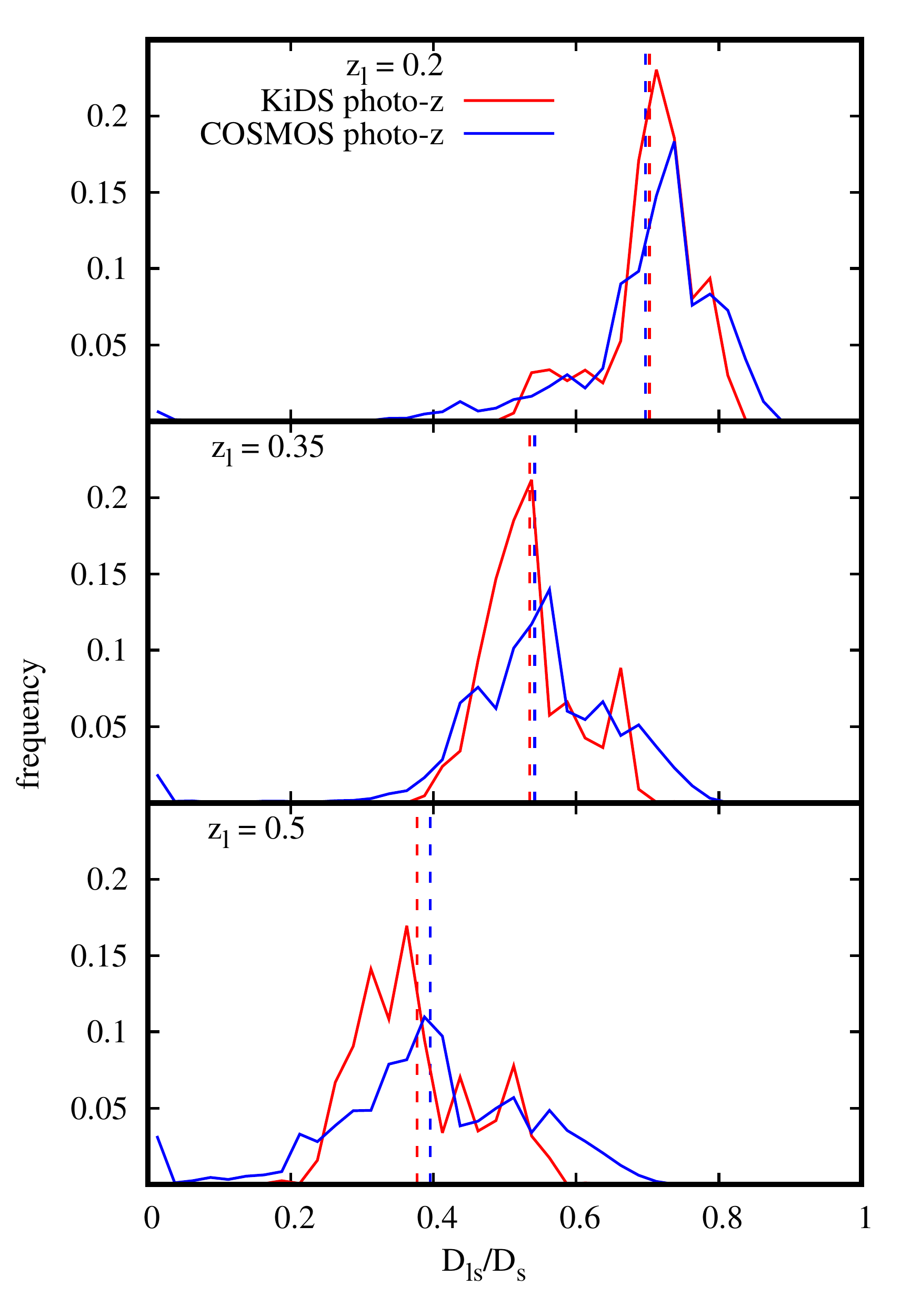}
 \caption{Left: redshift distributions of the background sample, as estimated from photo-$z$s by KiDS (red line) and COSMOS (blue line). The black dashed line represents the assumed cluster redshift, i.e. 0.2, 0.35 and 0.5 for the top, middle and bottom panel, respectively. Right: distribution of $D_\text{ls}/D_\text{s}$ for the background sample, as estimated from KiDS photometric redshifts (red line) and COSMOS ones (blue line). The dashed lines represent the mean $D_\text{ls}/D_\text{s}$ for the sample. The top, middle and bottom panels refer to different cluster redshifts, with $z_\text{l}$ = 0.2, 0.35 and 0.5, respectively.}
 \label{fig:comp_cosmos}
\end{figure*}

The method presented before aims at removing the selection effects from the considered spec-$z$ distribution. A complementary technique is to make use of external high-quality photometric redshifts for a complete subsample of galaxies. We can take advantage of the overlap of one tile in KiDS DR3 with the COSMOS field, where \citet{2016ApJS..224...24L} derived photometric redshifts for half a million galaxies with a precision $\sigma_z = 0.007 \times (1+z)$. The catalog is 98.6\% complete for $i^+ < 25$ (see their Figure 8), which is $\sim$ 1 mag deeper than KiDS. We can thus assume that virtually each object in the KiDS catalogue has a counterpart in the COSMOS catalogue. In this way, we do not need to weight COSMOS galaxies, as done by \citet{2018arXiv180500039M} and \citet{2018arXiv180405873M}. The results of this Section will depend on one KiDS tile only, but we do not introduce any uncertainty related to the weighting scheme. This test nicely complements the previous one, where we weighted galaxies from a spectroscopic subsample against the full KiDS catalogue. 

\begin{figure}
\includegraphics[width=\columnwidth]{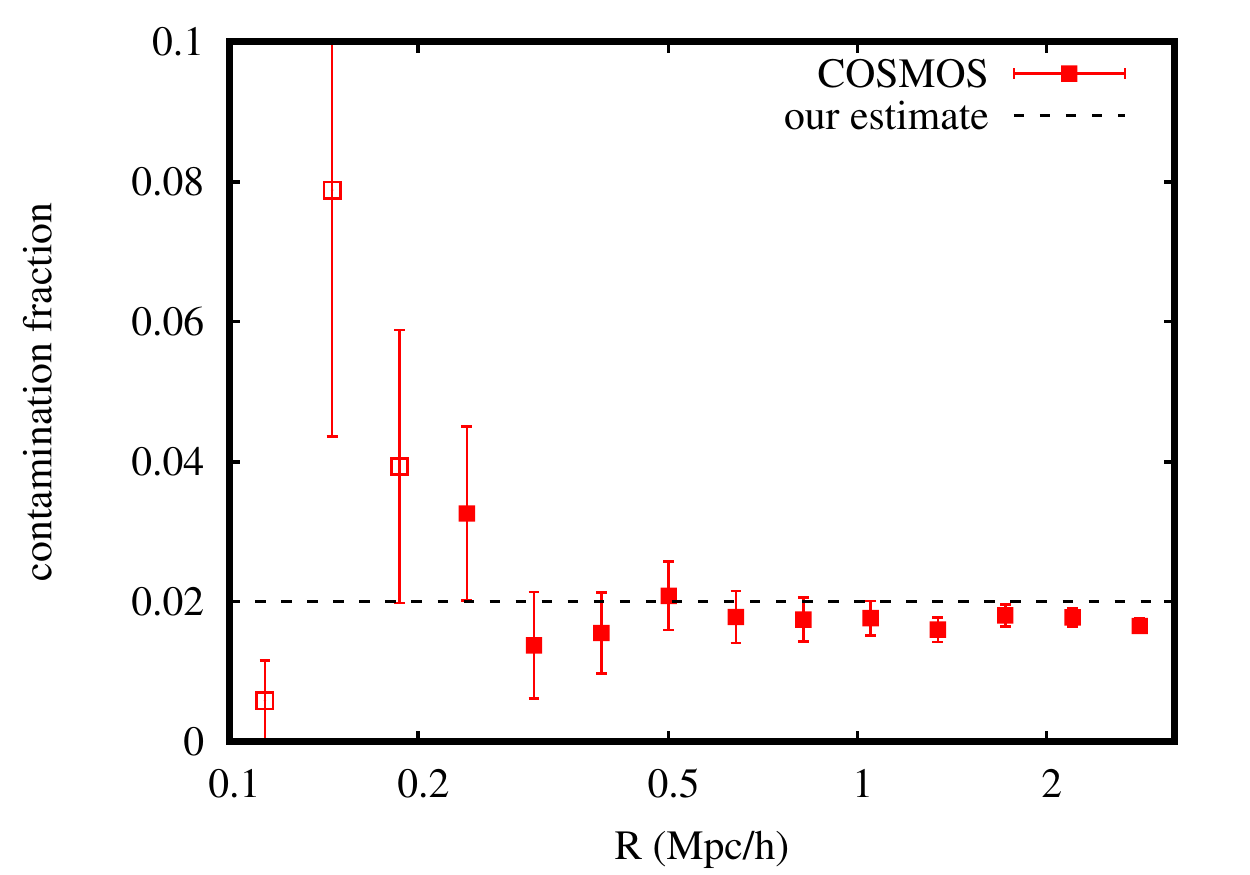}
 \caption{Fraction of background contaminants as a function of radius, as estimated from the COSMOS field. Empty points indicate radial bins not used in the lensing analysis. The dashed black line represents the preliminary estimate of Section \ref{sect:syst_prel}.}
 \label{fig:cont_cosmos}
\end{figure}

In practice, for each KiDS background galaxy, we search for COSMOS objects inside a radius of 1$\arcsec$. In the rare case of a double match, we choose the COSMOS galaxy with the lowest difference in $r$-band magnitude with respect to KiDS. In this way, we obtain a counterpart for the 99\% of the objects. We can then perform the same analysis we did in Section \ref{sect:comp_spec}, again considering three typical cluster redshifts 0.2, 0.35 and 0.5. The comparison between the KiDS and COSMOS background redshift distribution is shown in the left panel of Figure \ref{fig:comp_cosmos}. We underline that, since in this case we are considering a single KiDS tile, the intrinsic redshift distribution may be different from the one shown in Figure \ref{fig:comp_spec}. In fact, the distribution is slightly shifted towards higher redshifts. The shape of the COSMOS distribution resembles the KiDS one, but shows a tail at $z$ > 1.5 which is missing in our data and in the spec-$z$ distribution shown in Figure \ref{fig:comp_spec}. The contamination fraction we derive from this analysis is 0.6\%, 1.8\% and 3.2\%, for the three lens redshifts respectively, in remarkable agreement with the one derived in the previous Section and with our preliminary estimation. We can then translate this redshift distribution to a distribution in terms of $D_\text{ls}/D_\text{s}$. The results are shown in the right panel of Figure \ref{fig:comp_cosmos}. The resulting bias is 0.8\%, 1.3\% and 4.9\% for the three lens redshifts. Again, this is in agreement with the previous measurement with a spec-$z$ counterpart and with our analytic estimation.

As an additional test, we can estimate the contamination as a function of cluster radius, repeating the previous analysis only for galaxies that enter the background sample for at least one AMICO object. In this case, we consider as contaminants galaxies with $z_{\text{COSMOS}}$ < $z_{\text{l}} + 0.05$. The redshift buffer we adopt is larger than the typical uncertainty of cluster redshifts in the catalogue, that is $0.02(1+z_{\text{l}})$. This is likely to produce an overestimation of the contamination fraction as some background but close-by galaxies will be considered contaminants. As shown in Figure \ref{fig:cont_cosmos}, we do not detect any significant radial dependence, and we see a good agreement with our overall contamination estimation ($\sim 2$ per cent). Our conservative exclusion of the inner 200 kpc from the shear profile makes the effect of any residual radial dependence fully negligible. This confirms that the additional contamination due to cluster galaxies in the radial range of interest is compatible with our expectations and taken into account in our systematic uncertainty.
\end{subsubsection}

\end{subsection}

\begin{subsection}{Shear measurement}\label{sect:syst_shear}
According to Appendix D3 in \citet{2017MNRAS.465.1454H}, the uncertainty on the residual multiplicative bias in the shear estimates of KiDS DR3 is 1\%. This is an improvement of a factor 3 with respect to the previous Data Release. This uncertainty on the shear estimate of each galaxy, divided by $\Gamma$ (see Equation \ref{eq:gamma}), produces an uncertainty on mass of $\sim$ 1.3\%, which is sub-dominant with respect of those related to the selection of background galaxies and the estimate of their redshifts. We note that the shear responsivities have not been calibrated as a function of cluster radius, differently from e.g. \citet{2018arXiv180405873M}.
\end{subsection}

\begin{subsection}{Analytical modelling}\label{sect:syst_model}
If the stacked surface density profile does not follow the assumed model, systematic biases will be introduced in the mass retrieved from shear profiles. For what concerns the main halo, for masses $M_{200}<5\times10^{14}/h$, the expected error due to the choice of the model is $\lesssim 1$ per cent \citep{2016JCAP...01..042S}. At larger scales, the model presented in Section \ref{sect:model} has been shown in \citet{2011MNRAS.414.1851O} to describe the surface density distribution of cluster-sized halos in simulations better than non-truncated models. In particular, the model we use is effective in describing the transition region between the main halo and the correlated structures. \citet{2011MNRAS.414.1851O} show that using a simple NFW profile to fit a shear signal on these scales can produce biases up to 10-15\%, depending on the maximum radius considered. Thus, by truncating the main halo and by considering the contribution from correlated structures, we alleviate the possible mismatch between the real halo density profile and the model. For comparisons between these models and real data, see e.g. \citet{2013MNRAS.436.2616B} and \citet{2014ApJ...795..163U}. 

We evaluate the residual uncertainty related to halo modelling by measuring how much the retrieved masses are sensible to the precise definition of model details. Different prescriptions exist for the halo bias $b_h$, which determines the 2-halo term (Equation \ref{eq:2-halo}). We test the sensitivity of our results on $b_h$ by increasing the assumed values from \citet{2010ApJ...724..878T} by 10\%. We obtain a decrease in mass which depends on mass and redshift, typically $\sim$ 1\%.  A similar test can be performed on the truncation radius $r_\text{t}$, which we kept fixed at $r_\text{t} = 3 r_{200}$ during the analysis. We can vary it between 2.5$r_{200}$ and 4$r_{200}$, which is roughly the range of best-fitting values in \citet{2011MNRAS.414.1851O}, depending on mass and redshift. \footnote{\citet{2011MNRAS.414.1851O} quote their truncation parameter in terms of $r_{\text{vir}}$ instead of $r_{200}$.} We find a systematic increase (decrease) in mass for lower (higher) values of $r_\text{t}$, of the order of 1\%. Considering these tests, and residual percent-level discrepancies between the model and simulations shown in \citet{2011MNRAS.414.1851O} around $r_{\text{vir}}$ (see their Figure 3), we assume a 3\% systematic uncertainty due to halo modelling.

\end{subsection}

\begin{subsection}{Orientation and projections}\label{sect:syst_triax}
The halo model we use assumes spherical symmetry. As real halos are typically ellipsoidal \citep{2018ApJ...860L...4S}, the model is strictly valid only if the orientation of the halos in a stack is random. Moreover, the 2-halo term assumes a random distribution of correlated structures. Both these assumptions may not be accurate when dealing with optically-selected cluster samples.

\citet{2014MNRAS.443.1713D} tested different optical cluster finders on simulations and verified that they tend to select clusters with the main axis along the line of sight. This is expected, as they produce a larger observed galaxy number density contrast with respect to the field distribution. The resulting stacked mass distribution can be modelled as a prolate ellipsoid with the axis along the line of sight elongated by a factor $q$ $\sim$ 1.1. This result holds for all cluster finders tested, which use different methodologies and observables. Because of that, we can then safely assume the same for clusters detected in \citet{2018arXiv181002811M} and analysed in this work, even if AMICO was not tested in \citet{2014MNRAS.443.1713D}.  As lensing is sensitive to the projected mass over the plane of the sky, masses of halos elongated along the line of sight are biased high. \citet{2014MNRAS.443.1713D} derive this bias as $4.5 \pm 1.5$ \% for $q$ $\sim$ 1.1, with the uncertainty related to concentration of the halo.

Secondary halos aligned with the detected cluster can affect the mass estimate too. In fact, the observed amplitude of a cluster may be increased by the presence of other structures along the same line of sight, which are not detected and therefore will be blended in a single detection. Understanding how this effect influences the recovered masses is not trivial. For an approximate but effective estimation, we follow \citet{2017MNRAS.466.3103S} and \citet{2017MNRAS.469.4899M}, where the bias in the mass measurement of a stacked sample has been estimated as
\begin{equation}\label{eq:proj}
\frac {\delta M}{M} = \frac {p(\epsilon - 0.5)}{1-p(\epsilon - 0.5)} .
\end{equation}
Here, $p$ is the fraction of halos which suffer from projections and $\epsilon$ characterises the effective mass contribution of the projected halo. \citet{2017MNRAS.466.3103S} set $\epsilon = 0.25 \pm 0.15$ to cover a realistic range of possible contributions to the weak-lensing mass by projected haloes. To estimate $p$ for our sample, we note that, as an effect of blending, we expect to miss some objects which are aligned with the detected ones, as verified with simulations in \citet{2018MNRAS.473.5221B}. The projection ratio $p$ can then be estimated as the fraction of expected aligned detections which we are missing. To this aim, we consider for each object in our sample a cylinder with height equal to 3$\sigma_z$ and radius equal to 1.5 times the assumed $R_{200}$ of our model \citep[see][]{2018arXiv181002811M}. Comparing the density of detections inside these cylinders to the average one, we find a decrement of 15\%. We can then set this as the fraction of clusters $p$ which suffer from significant projections. Inserting this value in Equation \ref{eq:proj}, the resulting mass bias is $-0.04 \pm 0.02$  \%.

As projection effects due to blending of aligned structures counterbalance those due to the triaxial structure of the main halo, mass estimates do not need to be corrected. The residual uncertainty is taken into account by adding a systematic error of 3\% to our budget.

\end{subsection}
\begin{subsection}{Final assessment of systematics}

\begin{table}
\centering	
\caption{Sources of systematic uncertainty in the mass calibration described in Section \ref{sect:syst_unc}.}
\label{tab:syst_unc}
\begin{tabular}{ccc}
\hline
Source & $\Delta \Sigma$ error [\%]& Mass error [\%]\\
\hline
Background selection & 2 & 2.7\\
Photo-$z$ uncertainty & 4.2 & 5.6\\
Shear measurement & 1 & 1.3\\
Halo model & - & 3\\
Orientation and projections & - & 3\\
\hline
Total & 4.8 & 7.6\\
\hline
	\end{tabular}
\end{table}

The comparisons performed in Sections \ref{sect:comp_spec} and \ref{sect:comp_cosmos}, with different methodologies and assumptions, confirmed our preliminary estimates of Section \ref{sect:syst_prel}. Thus, we use those values in the following as our systematic uncertainties. We also consider the terms related to shear estimation discussed in Section \ref{sect:syst_shear}, the halo modelling (Section \ref{sect:syst_model}) and the halo triaxiality and projections (Section \ref{sect:syst_triax}). The adopted systematics are summarised in Table \ref{tab:syst_unc}. The total systematic uncertainty on the cluster masses is 7.6\%. We note that here we do not treat systematics related to concentration or miscentring because we include them in the model and thus mass posteriors naturally contain these sources of uncertainty.

 In the Appendixes we show the results of additional tests we performed to check the consistency of our results. In particular, we show the comparison between the signal extracted from the two selection criteria of background sources (App. \ref{sect:app_backsel}), the cross-correlation between background sources and cluster lenses (App. \ref{sect:app_cc}) and two null tests: the lensing signal around random positions and the radial profile of the cross-component of the shear (App. \ref{sect:app_null}).
\end{subsection}

\end{section}

\begin{section}{Mass-observable relation}\label{sect:mass-obs}
We model the relation between the photometric observable $O$ and the weak lensing mass as
\begin{equation}\label{eq:mass-obs}
\log {\frac {M_{200}}{10^{14} M_\odot/h}} = \alpha + \beta \log {\frac{O}{O_{\text{piv}}}}  + \gamma \log {\frac {E(z)}{E(z_{\text{piv}})}} ,
\end{equation}
where $E(z) = H(z)/H_0$ and $O_{\text{piv}}$ and $z_{\text{piv}}$ represent typical values of observable and redshift of the total sample, respectively. The redshift evolution of the relation is accounted for by the factor $\gamma \log E(z)$, following the approach by \citet{2015MNRAS.450.3675S}. For each bin of clusters, we compute the typical value of the observable $O_K$ through a lensing-weighted average \citep{2014ApJ...795..163U}
\begin{equation}\label{eq:bin_prop}
O_K = \frac {\sum_{k \in K} W_k O_k}{\sum_{k \in K} W_k},
\end{equation}
where $W_k$ is the total weight for the the $k$-th cluster of the bin,
\begin{equation}
W_k = \sum_i w_i \Sigma_{\text{crit},i}^{-2}
\end{equation}
and $i$ runs over all the background galaxies of the $k$-th cluster, irrespective of the radial bin. We follow the same procedure to calculate the typical redshift $z_K$ of each bin.

We build the covariance matrix $C_\text{M}$ for the mass estimates of the cluster bins as
\begin{equation}\label{eq:scal_rel_cov_matr}
C_{\text{M},ij} = \delta_{ij} E_i^2 + S^2 M_{200,i} M_{200,j},
\end{equation}
where $E_i$ represents the statistical error derived from the posterior distribution of $M_{200}$, and $S$ is the systematic uncertainty derived from the analysis in Section \ref{sect:syst_unc}. The systematic errors are treated as correlated, as they affect in a similar manner all cluster bins. In practice, we sum in quadrature the uncertainties for background selection, photo-$z$s and shear measurement, and we obtain $S$ = 0.076. This uncertainty is anyway sub-dominant with respect to the one derived from the $M_{200}$ posterior for all the bins. The systematic error instead dominates the uncertainty on the intercept of the mass-observable relation, as we will see in the next Section.

The parameters $\alpha$, $\beta$ and $\gamma$ of the mass-observable relation of Equation \ref{eq:mass-obs} are obtained with a Bayesian analysis analogous to the one presented in Section \ref{sect:der_par}, where we derived the parameters describing the lens matter distribution for each cluster bin. The likelihood is again given by Equation \ref{eq:like} where $\chi^2$ is now
\begin{equation}
\chi^2 = \sum_{i=1}^{N_{\text{bin}}} \sum_{j=1}^{N_{\text{bin}}} [M_{\text{obs},i} - M_{\text{mod},i}]\, \vec C^{-1}_{M,ij}\, [M_{\text{obs},j} - M_{\text{mod},j}]  .
\end{equation}
In the previous equation, the sums run over the $N_{\text{bin}}$ cluster bins, $M_{\text{obs}}$ is the measured $M_{200}$, $M_{\text{mod}}$ is the $M_{200}$ derived from the scaling relation and $C_{\text{M},ij}$ is given by Equation \ref{eq:scal_rel_cov_matr}. The priors are uniform with ranges given by $-1 < \alpha < 1$, $0.1 < \beta < 5$ and $-5 < \gamma < 5$. Also in this case, we used \textsc{MultiNest} for the derivation of the parameters and the quoted results are the mean and the standard deviation of the marginal posterior distribution.

\end{section}

\begin{section}{Results}\label{sect:results}
\subsection{Mass-amplitude relation}\label{sect:ampl-mass}

\begin{figure*}
 \includegraphics[width=2\columnwidth]{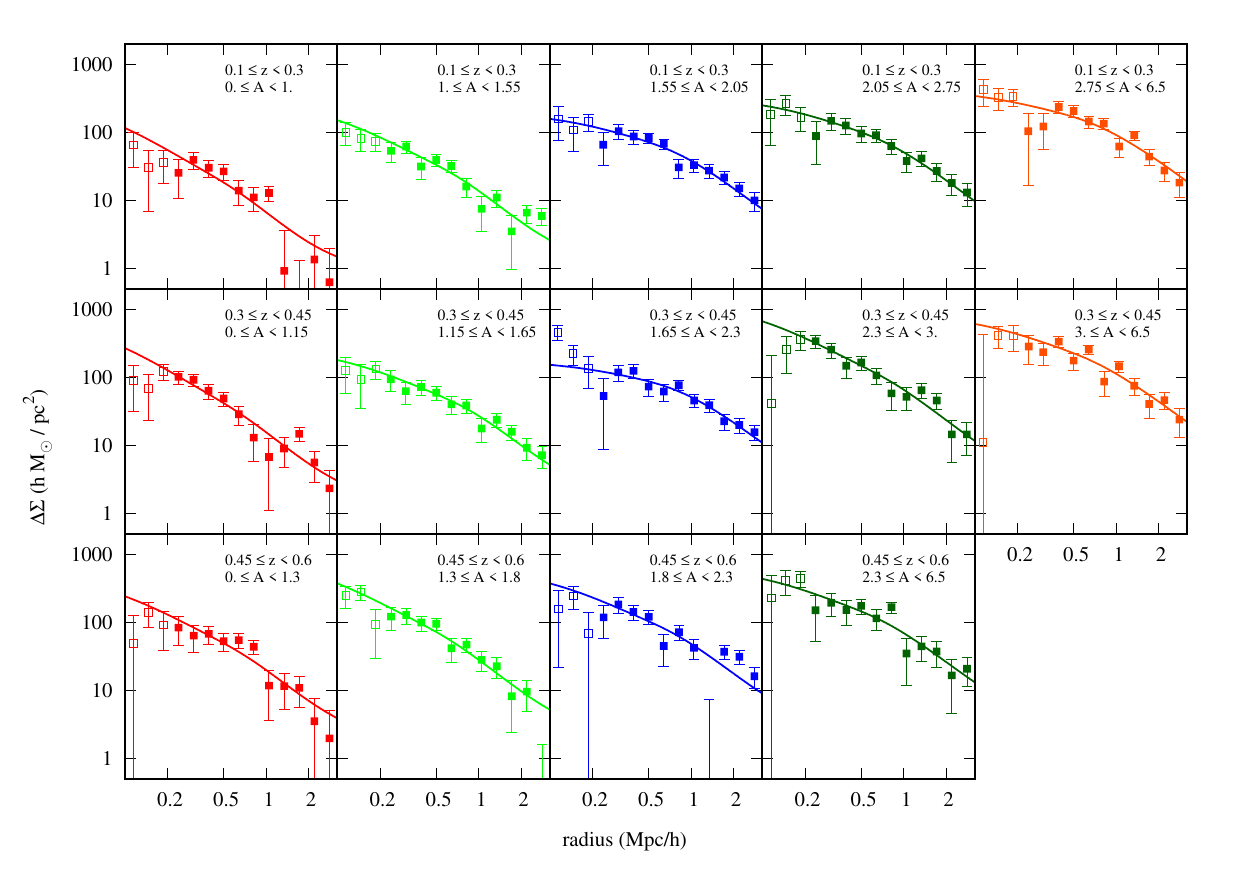}
 \caption{Stacked profiles for $\Delta \Sigma (R)$ in different bins of amplitude and redshift. Redshift increases from top to bottom, amplitude increases from left to right. Details of the properties for each bin can be found in Table \ref{tab:stack_bins}. In each box, the curve represents the model with mean posterior parameters. Empty squares represent data inside 0.2 \text{Mpc}/$h$ neglected in the fitting procedure.}
 \label{fig:shear_prof_array}
\end{figure*}

\begin{table*}
\centering	
\caption{Cluster binning used for the analysis presented in Section \ref{sect:ampl-mass}. The typical quantities $z_\text{eff}$ and $A_\text{eff}$ are computed according to Equation \ref{eq:bin_prop}. For $M_{200}$, we quote the mean and the standard deviation of the posterior probability distribution. $N_\text{cl}$ is the number of clusters in the bin. The number of degrees of freedom for $\chi^2$ comparison is 11 (data points) - 4 (parameters) = 7.}
\label{tab:stack_bins}
\begin{tabular}{ccccccc}
\hline
$z$ range & $z_{\text{eff}}$ & $A$ range & $A_{\text{eff}}$ & $M_{200} (10^{14} M_\odot/h)$ & $N_\text{cl}$ & $\chi^2$ \\
\hline
$[0.10,0.30[$ & 0.190 $\pm$ 0.002 & $[0,1[$ & 0.830 $\pm$ 0.003 & 0.145 $\pm$ 0.029 & 1066 & 12.4\\
$[0.10,0.30[$ & 0.207 $\pm$ 0.002 & $[1,1.55[$ & 1.213 $\pm$ 0.006 & 0.325 $\pm$ 0.054 & 822 & 11.3\\
$[0.10,0.30[$ & 0.212 $\pm$ 0.004 & $[1.55,2.05[$ & 1.762 $\pm$ 0.010 & 1.091 $\pm$ 0.144 & 240 & 6.0\\
$[0.10,0.30[$ & 0.226 $\pm$ 0.005 & $[2.05,2.75[$ & 2.350 $\pm$ 0.021 & 1.584 $\pm$ 0.233 & 96 & 2.5\\
$[0.10,0.30[$ & 0.211 $\pm$ 0.008 & $[2.75,6.5[$ & 3.259 $\pm$ 0.084 & 3.450 $\pm$ 0.429 & 41 & 11.7\\
\hline
$[0.30,0.45[$ & 0.378 $\pm$ 0.001 & $[0,1.15[$ & 0.954 $\pm$ 0.004 & 0.376 $\pm$ 0.061 & 1090 & 9.6\\
$[0.30,0.45[$ & 0.382 $\pm$ 0.002 & $[1.15,1.65[$ & 1.354 $\pm$ 0.006 & 0.686 $\pm$ 0.115 & 762 & 3.6\\
$[0.30,0.45[$ & 0.385 $\pm$ 0.002 & $[1.65,2.3[$ & 1.909 $\pm$ 0.012 & 1.485 $\pm$ 0.188 & 339 & 6.1\\
$[0.30,0.45[$ & 0.392 $\pm$ 0.004 & $[2.3,3[$ & 2.585 $\pm$ 0.022 & 2.079 $\pm$ 0.360 & 98 & 5.4\\
$[0.30,0.45[$ & 0.377 $\pm$ 0.007 & $[3,6.5[$ & 3.577 $\pm$ 0.071 & 4.114 $\pm$ 0.558 & 43 & 12.0\\
\hline
$[0.45,0.60[$ & 0.496 $\pm$ 0.001 & $[0,1.3[$ & 1.108 $\pm$ 0.004 & 0.469 $\pm$ 0.087 & 984 & 6.3\\
$[0.45,0.60[$ & 0.516 $\pm$ 0.002 & $[1.3,1.8[$ & 1.516 $\pm$ 0.005 & 0.711 $\pm$ 0.114 & 889 & 8.8\\
$[0.45,0.60[$ & 0.515 $\pm$ 0.003 & $[1.8,2.3[$ & 2.071 $\pm$ 0.011 & 1.358 $\pm$ 0.244 & 373 & 23.3\\
$[0.45,0.60[$ & 0.510 $\pm$ 0.004 & $[2.3,6.5[$ & 2.877 $\pm$ 0.039 & 2.120 $\pm$ 0.396 & 119 & 9.3\\
\hline

	\end{tabular}
\end{table*}

\begin{figure}
 \includegraphics[width=\columnwidth]{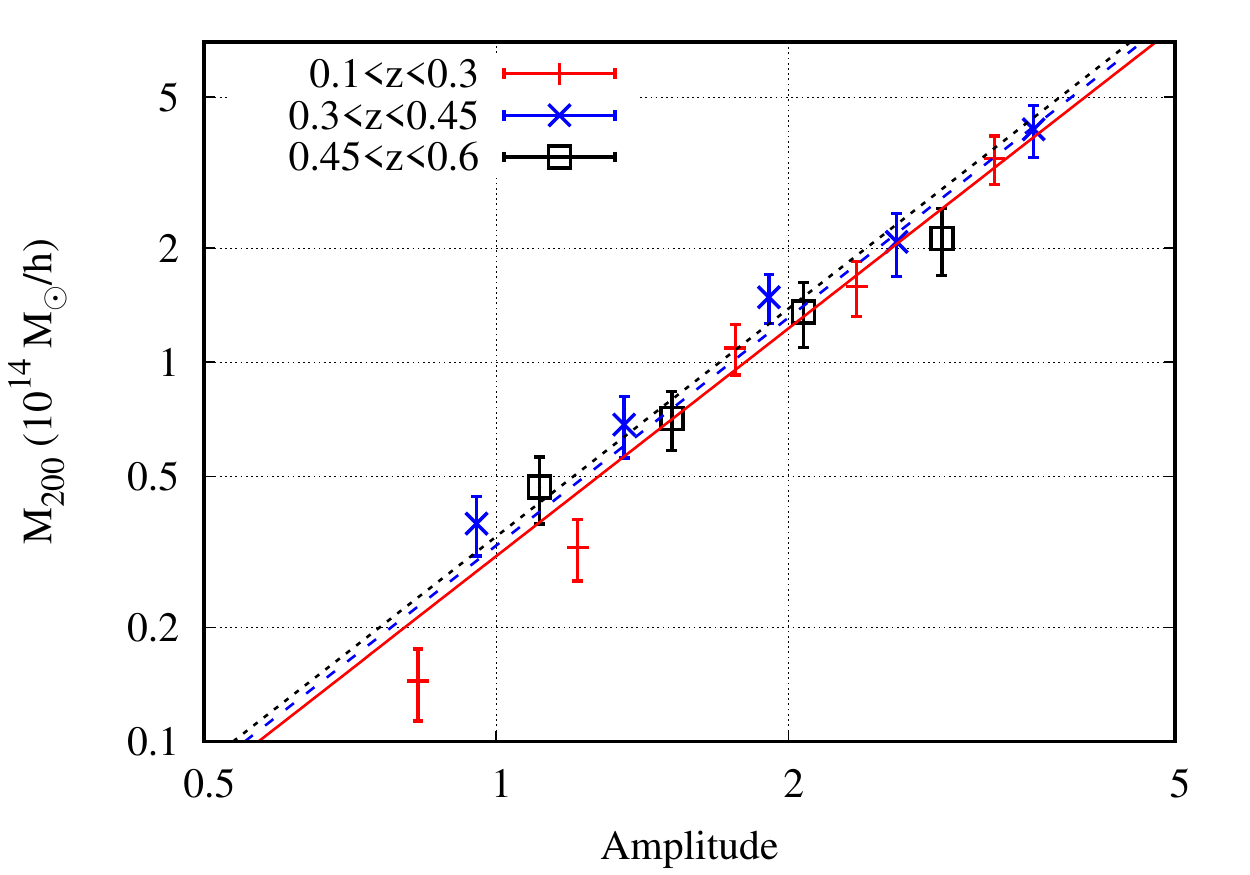}
 \caption{Weak lensing mass of the stacked shear profile as a function of amplitude, in different redshift bins. The vertical error bars represents the errors derived from the posterior distribution plus the systematic uncertainties. The mass-amplitude relation (Equation \ref{eq:mass-obs}) is shown for the typical redshift of each redshift bin (0.21, 0.37, 0.50).}
 \label{fig:stack_amp_mass}
\end{figure}

We divide the cluster sample presented in \ref{sect:clu_cat} in three redshift intervals: 0.1 $\leq z$ < 0.3, 0.3 $\leq z$ < 0.45, 0.45 $\leq z$ < 0.6. Each of them is then additionally subdivided in amplitude, creating 5, 5 and 4 amplitude bins respectively, for a total of 14 redshift-amplitude bins. For each bin, we extract the stacked shear profile and we derive a mass estimate from it, as detailed in Sections \ref{sect:profile}, \ref{sect:model} and \ref{sect:der_par}. The results are shown in Figure \ref{fig:shear_prof_array} and reported in Table \ref{tab:stack_bins}. We can see there is a clear correlation between excess surface density and amplitude in each redshift range. Typically, the most likely model describes well the data in the radial range $0.2 < R < 3$ Mpc/$h$ used in the analysis, as confirmed by the $\chi^2$ values shown in Table \ref{tab:stack_bins}.

We then use the masses derived for each amplitude-redshift cluster bin to constrain the mass-amplitude relation (Equation \ref{eq:mass-obs}). We set $z_{\text{piv}}$ = 0.35 and $A_{\text{piv}}$ = 2, as these are central values in the ranges covered by the whole sample. Following the analysis detailed in Section \ref{sect:mass-obs}, we derive the parameters describing the intercept $\alpha$, the slope $\beta$ and the redshift evolution $\gamma$ of the mass-amplitude relation. The results are:
\begin{itemize}
\item $\alpha =  0.114 \pm 0.038$
\item  $\beta = 1.99 \pm 0.10$
\item $\gamma = 0.73 \pm 0.63$
\end{itemize}
The resulting relation for three typical redshift values ($z = 0.21, 0.37, 0.50$) is shown in Figure \ref{fig:stack_amp_mass}, together with the mass-amplitude data points derived from the analysis. 

The mass-amplitude relation is linear across the redshift and amplitude range covered by the sample. It is remarkable that the same mass-amplitude relation fits the data well for more than one order of magnitude in mass, down to $M_{200}$ $\sim$ 2 (5)  $\times 10^{13} M_\odot/h$ at $z$ = 0.2 (0.5). There is no sign that clusters in different amplitude regimes follow different relations, that could come, for instance, from fake detections contaminating the low $A$ samples. The slope of the mass-amplitude relation is compatible with the slope of the inverse relation (amplitude-mass) $\sim 0.54 \pm 0.04$ derived from simulations in \citet{2018MNRAS.473.5221B}. The relation is compatible with being time-independent at the $\sim 1 \sigma$ level.

We note that systematic errors dominate in determining the uncertainties in the intercept $\alpha$, while being negligible for $\beta$ and $\gamma$.

\subsection{Mass-richness relation}\label{sect:lambda-mass}

\begin{figure*}
 \includegraphics[width=2\columnwidth]{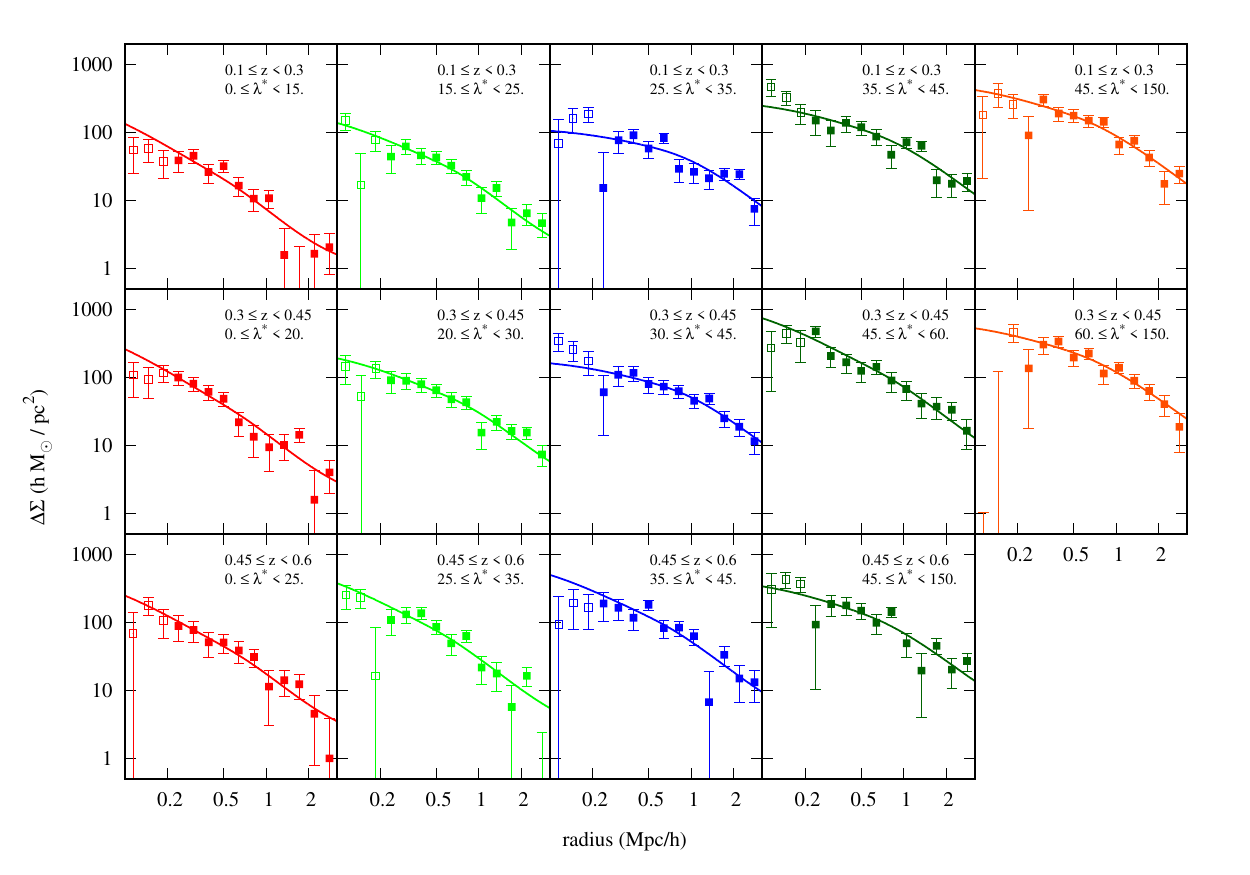}
 \caption{Stacked profiles for $\Delta \Sigma (R)$ in different bins of richness $\lambda^\text{\textasteriskcentered}$ and redshift. Redshift increases from top to bottom, richness increases from left to right. Details of the properties for each bin can be found in Table \ref{tab:stack_bins_lambda}. In each box, the curve represents the model with mean posterior parameters. Empty squares represent data inside 0.2 \text{Mpc}/$h$ neglected in the fitting procedure.}
 \label{fig:shear_prof_array_lambda}
\end{figure*}

\begin{table*}
\centering	
\caption{Cluster binning used for the analysis presented in Section \ref{sect:lambda-mass}. The typical quantities $z_\text{eff}$ and $\lambda^\text{\textasteriskcentered}_\text{eff}$ are computed according to Equation \ref{eq:bin_prop}. For $M_{200}$, we quote the mean and the standard deviation of the posterior probability distribution. $N_\text{cl}$ is the number of clusters in the bin. The number of degrees of freedom for $\chi^2$ comparison is 11 (data points) - 4 (parameters) = 7. }
\label{tab:stack_bins_lambda}
\begin{tabular}{ccccccc}
\hline
\hline
$z$ range & $z_{\text{eff}}$ & $\lambda^\text{\textasteriskcentered}$ range & $\lambda^\text{\textasteriskcentered}{\text{eff}}$ & $M_{200} (10^{14} M_\odot/h)$ & $N_\text{cl}$ & $\chi^2$ \\
\hline
$[0.10,0.30[$ & 0.189 $\pm$ 0.001 & $[0,15[$ & 10.20 $\pm$ 0.09 & 0.165 $\pm$ 0.026 & 1246 & 9.8\\
$[0.10,0.30[$ & 0.212 $\pm$ 0.002 & $[15,25[$ & 18.88 $\pm$ 0.12 & 0.383 $\pm$ 0.062 & 684 & 5.5\\
$[0.10,0.30[$ & 0.223 $\pm$ 0.004 & $[25,35[$ & 29.02 $\pm$ 0.21 & 1.070 $\pm$ 0.163 & 209 & 20.1\\
$[0.10,0.30[$ & 0.228 $\pm$ 0.007 & $[35,45[$ & 39.75 $\pm$ 0.32 & 2.027 $\pm$ 0.297 & 82 & 10.1\\
$[0.10,0.30[$ & 0.222 $\pm$ 0.008 & $[45,150[$ & 56.59 $\pm$ 2.20 & 3.222 $\pm$ 0.413 & 44 & 11.0\\
\hline
$[0.30,0.45[$ & 0.374 $\pm$ 0.001 & $[0,20[$ & 15.10 $\pm$ 0.11 & 0.355 $\pm$ 0.056 & 1113 & 10.0\\
$[0.30,0.45[$ & 0.388 $\pm$ 0.002 & $[20,30[$ & 24.08 $\pm$ 0.11 & 0.788 $\pm$ 0.112 & 767 & 7.1\\
$[0.30,0.45[$ & 0.389 $\pm$ 0.002 & $[30,45[$ & 35.91 $\pm$ 0.27 & 1.509 $\pm$ 0.210 & 320 & 4.0\\
$[0.30,0.45[$ & 0.390 $\pm$ 0.005 & $[45,60[$ & 50.88 $\pm$ 0.50 & 2.341 $\pm$ 0.413 & 87 & 5.3\\
$[0.30,0.45[$ & 0.379 $\pm$ 0.006 & $[60,150[$ & 73.60 $\pm$ 2.09 & 4.487 $\pm$ 0.605 & 45 & 7.8\\
\hline
$[0.45,0.60[$ & 0.498 $\pm$ 0.001 & $[0,25[$ & 19.71 $\pm$ 0.11 & 0.410 $\pm$ 0.084 & 1108 & 3.5\\
$[0.45,0.60[$ & 0.514 $\pm$ 0.002 & $[25,35[$ & 29.23 $\pm$ 0.12 & 0.759 $\pm$ 0.115 & 761 & 14.3\\
$[0.45,0.60[$ & 0.523 $\pm$ 0.003 & $[35,45[$ & 39.25 $\pm$ 0.18 & 1.475 $\pm$ 0.242 & 299 & 11.7\\
$[0.45,0.60[$ & 0.513 $\pm$ 0.004 & $[45,150[$ & 55.12 $\pm$ 0.76 & 2.119 $\pm$ 0.346 & 197 & 13.7\\
\hline

	\end{tabular}
\end{table*}

\begin{figure}
 \includegraphics[width=\columnwidth]{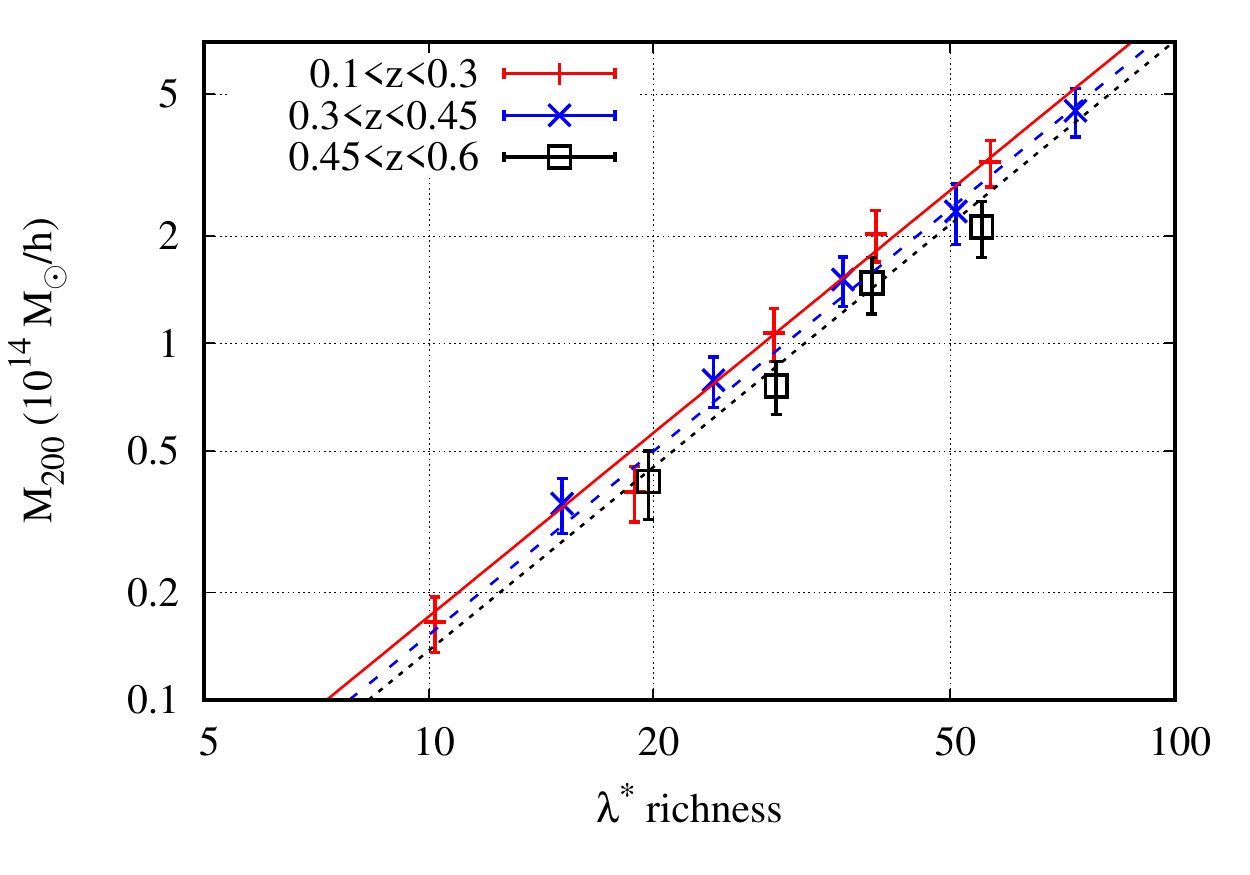}
 \caption{Weak lensing mass of the stacked shear profile as a function of richness $\lambda^\text{\textasteriskcentered}$, in different redshift bins. The vertical error bars represents the errors derived from the posterior distribution plus the systematic uncertainties. The mass-richness relation (Equation \ref{eq:mass-obs}) is shown for the typical redshift of each redshift bin (0.21, 0.37, 0.50).}
 \label{fig:stack_lambda_mass}
\end{figure}

AMICO assigns to each galaxy a probability of being part of any detected structure, according to Equation \ref{eq:member_prob_corr}. The tests performed in \citet[][see Fig. 8]{2018MNRAS.473.5221B} show that if the cluster model is a good description of the mean properties of the clusters in the data, the retrieved probability describes remarkably well the true statistical membership in the catalog. As a consequence, under the same assumption, the sum of membership probabilities for each structure will roughly correspond to the number of visible members. In principle, this quantity can be used as a mass proxy, but it is severely redshift-dependent, because the faint members will go beyond the survey limiting magnitude as the redshift increases. In order to remove the intrinsic redshift dependence, we need to consider a consistent sample of members, i.e. one that is observable over all the considered redshift range.

We then define the richness $\lambda^\text{\textasteriskcentered}_j$ of the $j$-th detection as
\begin{equation}\label{eq:lambda}
\lambda^\text{\textasteriskcentered}_j = \sum_{i=1}^{N_{gal}} P(i \in j) F_{ij}, 
\end{equation}
where $F_{ij}$ selects the galaxies according to
\begin{equation}\label{eq:lambda_cuts}
  F_{ij}=\begin{cases}
    1, & \text{ if } m_i < m_\text{\textasteriskcentered}(z_j)+1.5 \text{ and } R_i < R_{200}(z_j)\\
    0, & \text{otherwise}.
  \end{cases}
\end{equation}
and $P(i \in j)$ is given by Equation \ref{eq:member_prob_corr}. In the previous equation, $m_\text{\textasteriskcentered}$ and $R_{200}$ are parameters of the model presented in \citet{2018arXiv181002811M} and $z_j$ is the redshift of the detection. As will be discussed in Section \ref{sect:literature}, this definition of richness has similarities with those used by other cluster finders such as redMaPPer \citep{2014ApJ...785..104R} and RedGOLD \citep{2016MNRAS.455.3020L}. The upper limit at $m_\text{\textasteriskcentered}+1.5$ is brighter than the magnitude limit of the survey at all redshifts covered by clusters in this work. In this way, we define a $\lambda^\text{\textasteriskcentered}$ for each detection.

We then proceed similarly to Section \ref{sect:ampl-mass}, this time considering the richness $\lambda^\text{\textasteriskcentered}$ as the observable instead of the amplitude $A$. We thus bin the clusters according to their redshift and their $\lambda^\text{\textasteriskcentered}$, and we get again 14 redshift-richness bins. The resulting weak lensing profiles for each cluster bin and the corresponding weak lensing masses are given in Figure \ref{fig:shear_prof_array_lambda} and in Table \ref{tab:stack_bins_lambda}. The results are then used to constrain the mass-richness relation (Equation \ref{eq:mass-obs}), with $z_{\text{piv}}$ = 0.35 and $\lambda^\text{\textasteriskcentered}_{\text{piv}}$ = 30. This time, the resulting parameters are:
\begin{itemize}
\item $\alpha =  0.004 \pm 0.038$
\item $\beta = 1.71 \pm 0.08$
\item $\gamma = -1.33 \pm 0.64$.
\end{itemize}
The relation between richness and weak lensing mass is shown in Figure \ref{fig:stack_lambda_mass}.

As in the case of the mass-amplitude relation, the mass-richness linear relation holds well for all redshifts and all richnesses in the sample, down to the low mass limit of the cluster sample. The most notable difference with respect to the relation between amplitude and mass is that the slope $\beta$ is shallower (1.71 vs 1.99). Qualitatively, this corresponds to a steeper dependence of the richness on the mass in comparison to the amplitude. There is also a $\sim 2 \sigma$ trend towards a lower mass for higher redshift clusters at fixed richness, driven by the highest redshift bin. As in the mass-amplitude relation, the uncertainty on the intercept $\alpha$ is dominated by systematical uncertainties, while their effect is small on $\beta$ and $\gamma$.

As the catalogue of clusters has been selected in S/N (proportional to $A$) and not on $\lambda^\text{\textasteriskcentered}$, we note that selection effects may influence the lowest $\lambda^\text{\textasteriskcentered}$ bin at each redshift, i.e. objects with equivalent $\lambda^\text{\textasteriskcentered}$ but amplitude below the threshold did not enter the catalogue and are thus excluded from this derivation. As a consequence, the scaling relation we derived is strictly valid only for a sample of clusters selected according to AMICO criteria.
\section{Discussion}

\subsection{Comparison of the two observables}\label{sect:ampl-rich}

\begin{figure}
 \includegraphics[width=\columnwidth]{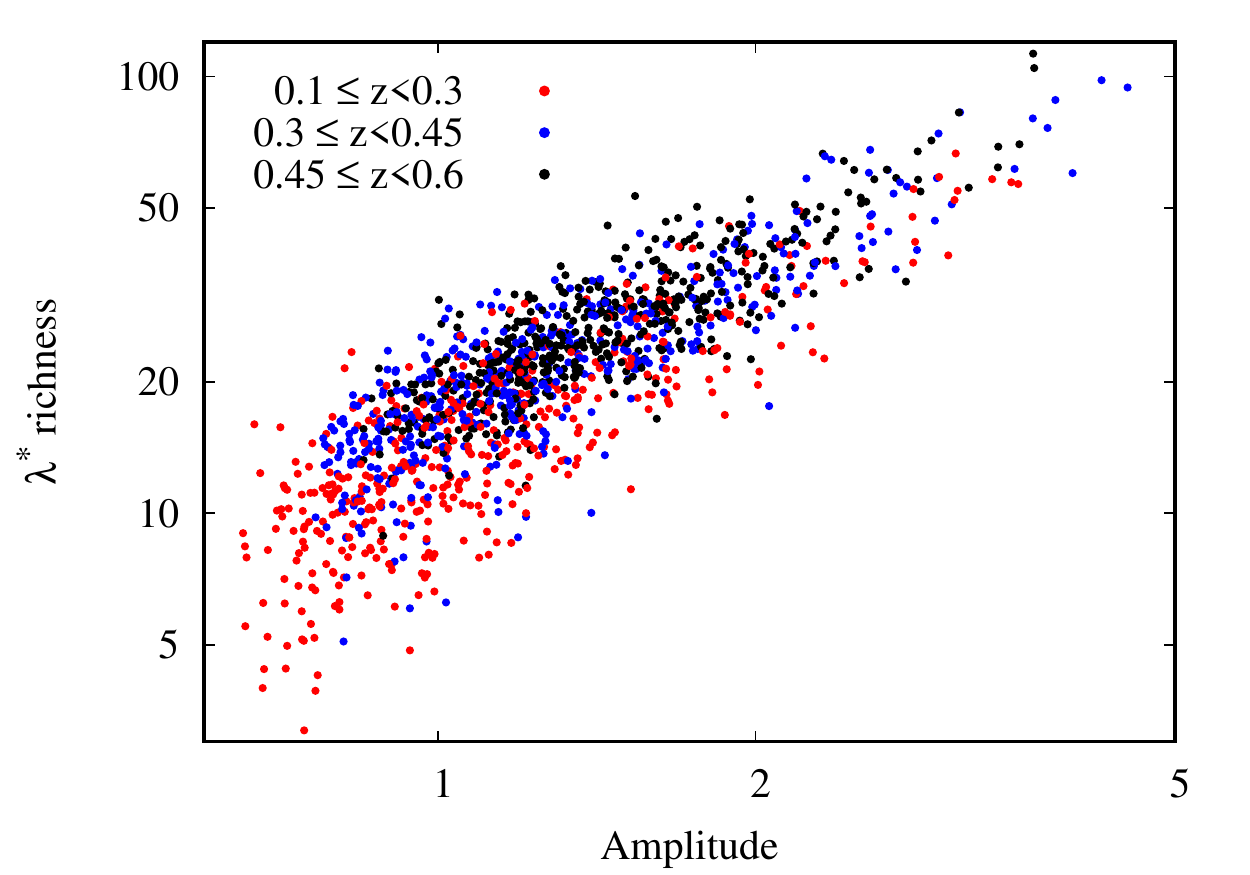}
 \caption{Relation between amplitude and $\lambda^\text{\textasteriskcentered}$ for the cluster sample. Point colours reflect the cluster redshift (0.1$\leq$z<0.3 in red, 0.3$\leq$z<0.45 in blue, 0.45$\leq$z<0.6 in black). To limit confusion, only one every 5 clusters is plotted.}
 \label{fig:amp-lam}
\end{figure}

The two observables analysed in Sections \ref{sect:ampl-mass} and \ref{sect:lambda-mass} are connected but different in nature. The amplitude $A$ is a measure of abundance of galaxies in units of the input cluster model. It is obtained by convolving the observed galaxy distribution with an optimal filter, which maximises the S/N of the detection. In its calculation, each galaxy is weighted by a function that depends on the ratio between the assumed cluster model and the background distribution (see Equation \ref{eq:amplitude}). The amplitude proved to be an unbiased mass proxy in simulations where the model describes the mean properties of the clusters and the galaxies' $p(z)$ follow a simple Gaussian distribution with a realistic dispersion \citep{2018MNRAS.473.5221B}. In this analysis on real data, we confirmed that the amplitude is an efficient mass proxy over all the considered redshift and mass ranges. 

The richness $\lambda^\text{\textasteriskcentered}$ defined in Section \ref{sect:lambda-mass} is the sum of membership probabilities inside redshift-independent ranges in magnitude and radius. It is clearly connected to the amplitude by construction through Equation \ref{eq:member_prob_corr}, and this is reflected in the clear correlation shown in Figure \ref{fig:amp-lam}. With respect to the amplitude $A$, the richness $\lambda^\text{\textasteriskcentered}$ has the advantage of being less dependent on the assumed cluster model. First of all, the membership probability is proportional to the product of $A$ and the model, which is independent of the input normalisation of the model. Then, the weight of each galaxy in $\lambda^\text{\textasteriskcentered}$ is by construction limited by 1, coherently with the definition of probability: this means that, while it is possible for the amplitude $A$ to be boosted by a small number of galaxies with an especially high weight, the effect of the same galaxies in the richness calculation would be significant but moderate. Finally, the values of $\lambda^\text{\textasteriskcentered}$ calculated according to Equations \ref{eq:lambda} and \ref{eq:lambda_cuts} can be reasonably extrapolated to other galaxy selections and compared to those extracted assuming other cluster models.

On the other hand, the amplitude $A$ is the primary observable of AMICO and has proven to be very effective in the detection procedure on mock and real data \citep{2018MNRAS.473.5221B,2018arXiv181002811M}, as it filters the data with an appropriate weight on a cluster scale. The selection function in terms of $A$ is simpler to model with respect to the one in $\lambda^\text{\textasteriskcentered}$, because the detection threshold in the cluster catalog is defined in terms of S/N = $A/\sigma_A$. This may be convenient especially if one wants to use the cluster catalogue and its calibration for cosmological purposes, when a precise knowledge of the selection function is mandatory.

\subsection{Comparison with literature}\label{sect:literature}
This is the first mass calibration of the cluster observables provided by AMICO, so there is no previous work to which we can directly compare our results. Not only the mass proxies we employed in this analysis have never been calibrated with different mass estimates on real data, but the very sample of clusters may have different properties with respect to the ones detected with other algorithms. Anyway, we can compare with previous works on three aspects: the properties of the observables, the slope of the mass-observable relation, and the mass range covered by the calibration.

\subsubsection{Properties of the observables}
The amplitude $A$ is derived by AMICO following Equation \ref{eq:amplitude}. The convolution with a cluster-sized kernel is typical of matched filters, but other cluster finders do not quote their result as an observable. For example, \citet{2015MNRAS.447.1304F}, when performing a calibration of the clusters detected with the 3D-MF algorithm \citep{2010MNRAS.406..673M}, used as observable an estimate of $N_{200}$ calculated a posteriori. We have shown that the AMICO amplitude is indeed a reliable mass proxy, providing the first mass calibration on real data of this kind of observable.

The richness $\lambda^\text{\textasteriskcentered}$ is more similar to known cluster observables in literature, as for example the richnesses calculated by cluster finding algorithms such as redMaPPer \citep{2014ApJ...785..104R} and RedGOLD \citep{2016MNRAS.455.3020L}. We share the idea of counting galaxies inside a given radius from the cluster centre, weighed by their probability of being cluster members. There are two main differences with respect to these works: we do not perform any cut based on the galaxy colour, instead of selecting red galaxies only, and we use a constant limiting radius for all clusters at the same redshift, instead of estimating the cluster $r_{200}$ on the same data. 


\subsubsection{Slope of the mass-observable relation}
The slopes we derived for the mass-amplitude relation ($1.99 \pm 0.10$) and the mass-richness relation ($1.71 \pm 0.08$) are significantly larger than typical results in literature about calibration of photometric observables. For example, the logarithmic dependence of the weak lensing mass on the richness $\lambda$ provided by redMaPPer has been found to be $1.12 \pm 0.26$ in \citet{2017MNRAS.469.4899M} and $1.33 ^{+0.09}_{-0.10}$ by \citet{2017MNRAS.466.3103S}, while \citet{2017ApJ...848..114P} derived a slope $1.02\pm0.21$ for the calibration of RedGOLD $\lambda$. This likely depends on the fact that AMICO uses a fixed radius in the definition of its observables, as already discussed in Section 5.3 of \citet{2018MNRAS.473.5221B}: clusters bigger than the model are cut by the kernel, smaller ones are slightly enhanced by correlated large-scale structure. In fact, a similarly steep slope is found in \citet{2017A&A...608A.141T}, which calibrates SpARCS clusters using as an observable $N_\text{red}$, the number of red galaxies in a fixed aperture (500 kpc). We note however that the slope of the mass-richness relation is flatter than the one of the mass-amplitude relation, indicating that this feature is less relevant for $\lambda^\text{\textasteriskcentered}$. This is expected, as the estimation of $\lambda^\text{\textasteriskcentered}$ is less dependent on the assumed model, and closer to the properties of the actual galaxy distribution in the data (see Section \ref{sect:ampl-rich}).


\subsubsection{Mass range of the calibration}
In this work we calibrated the cluster sample extracted by AMICO on KiDS DR3 data \citep{2018arXiv181002811M} down to a $M_{200}$ equal to $0.2 \times 10^{14} M_\odot/h$ at $z=0.2$ and to $0.5 \times 10^{14} M_\odot/h$ at $z=0.5$. This is competitive with the most up-to-date efforts for cluster detection at intermediate redshifts on photometric data. For comparison, \citet{2017MNRAS.469.4899M} shows the mass calibration of redMaPPer clusters detected on DES data at redshifts $0.2 \leq z \leq 0.8$. Their results are computed only on the high richness sample ($\lambda >$ 20), and extrapolate well down to $\lambda \ge$ 14, which corresponds to $M_{200} \sim 0.55 \times 10^{14} M_\odot/h$\footnote{In \citet{2017MNRAS.469.4899M} masses are given as $M_{200,\text{m}}$ in units of $M_\odot$. The conversion in terms of $M_{200,\text{c}}$ assumes $c_{200}$ = 4 and is roughly valid over 0.2 < $z$ < 0.6, the common redshift range among the two analyses.}.
 The catalog extracted from the RCS2 survey is calibrated by \citet{2016A&A...586A..43V} down to $\sim 0.4 \times 10^{14} M_\odot/h$ at $z \leq 0.8$ via a post-processing estimation of $N_{200}$. The RedGOLD sample detected on CFHTLS and NGVS is calibrated down to $M_{200} \sim 0.7 \times 10^{14} M_\odot/h$ \citep{2017ApJ...848..114P}.

\end{section}

\begin{section}{Summary and conclusions}
Mass calibration of optically-detected galaxy clusters is fundamental for their astrophysical and cosmological exploitation. In this work we derived, for the first time, weak lensing masses for clusters detected by AMICO on KiDS data \citep{2018arXiv181002811M}. The sample comprises 6962 galaxy clusters at redshifts between 0.1 and 0.6. Using KiDS shear data, we performed a stacked weak lensing analysis for ensembles of clusters selected according to their redshift and amplitude $A$, the optical mass proxy provided by AMICO. The amplitude is an estimate of the abundance of galaxies in units of an input cluster model, which was found to be a robust mass proxy on simulations in \citet{2018MNRAS.473.5221B}.

From a sample of background galaxies selected according to photo-$z$s and colours (see Section \ref{sect:sel_back}), we derived the stacked weak lensing signal for each bin in amplitude and redshift. We modelled the weak lensing signal as a truncated NFW distribution, plus a 2-halo term that describes the correlated matter around the cluster. In the analysis, we considered the radial region between 0.2 and 3 Mpc/$h$, to minimise uncertainties due to contribution by the BCG and the large-scale structure. The best parameters (and their errors) for each weak lensing profile were derived with a Bayesian Nested Sampling algorithm, \textsc{MultiNest}. The priors on concentration and fraction of miscentred halos were kept wide and conservative by choice, to marginalise over modelling uncertainties in the determination of the mean halo mass $M_{200}$. 

We found a clear weak lensing signal for all cluster bins in redshift and amplitude, and constructed a robust mass-amplitude relation. In its derivation, in addition to statistical errors, we took into account systematic uncertainties related to the possible contamination of the background sample, the photo-$z$ estimate of each galaxy and the shear measurement. The resulting relation between amplitude and mass is
\begin{equation}\label{eq:ampl-mass}
\log {\frac {M_{200}}{10^{14} M_\odot/h}} = \alpha + \beta \log {\frac{A}{A_{\text{piv}}}}  + \gamma \log {\frac {E(z)}{E(z_{\text{piv}})}} ,
\end{equation}
with parameters $\alpha =  0.114 \pm 0.038$, $\beta = 1.99 \pm 0.10$ and $\gamma = 0.73 \pm 0.63$, given $A_{\text{piv}}$ = 2 and $z_{\text{piv}}$ = 0.35. The slope $\beta$ is compatible to the results obtained by \citet{2018MNRAS.473.5221B} on simulations. The relation given by Equation \ref{eq:ampl-mass} holds for the whole redshift and amplitude ranges covered by the sample. 

In addition, we repeated the same analysis based on a second observable: the cluster richness $\lambda^\text{\textasteriskcentered}$, defined from the membership probability as provided by AMICO. In order to define a redshift-independent mass proxy, we selected members inside consistent ranges in radius and magnitude, specifically $R$ < $R_{200}$ and $m$ < $m_{\text{\textasteriskcentered}} + 1.5$, where $R_{200}$ and $m_\text{\textasteriskcentered}$ are redshift-dependent quantities defined according to our model \citep[see][]{2018arXiv181002811M}. The resulting mass-richness relation is given by
\begin{equation}\label{eq:rich-mass}
\log {\frac {M_{200}}{10^{14} M_\odot/h}} = \alpha + \beta \log {\frac{\lambda^\text{\textasteriskcentered}}{\lambda^\text{\textasteriskcentered}_{\text{piv}}}}  + \gamma \log {\frac {E(z)}{E(z_{\text{piv}})}} ,
\end{equation}
with parameters $\alpha =  0.004 \pm 0.038$, $\beta = 1.71 \pm 0.08$ and $\gamma = -1.33 \pm 0.64$, given $\lambda^\text{\textasteriskcentered}_{\text{piv}}$ = 30 and $z_{\text{piv}}$ = 0.35. The slope is flatter than the one for the mass-amplitude relation. The richness $\lambda^\text{\textasteriskcentered}$ should be less affected by limits in the modelling than $A$, and easier to compare among different surveys/models.

In this work we derived two consistent mass calibrations for the sample of clusters derived by AMICO on KiDS data and presented in \citet{2018arXiv181002811M}. These results confirm that AMICO is able to detect structures down to few $10^{13} M_\odot/h$ in the KiDS data, and to provide a reliable mass proxy for each of them. The application of the same recipes to future KiDS Data Releases will provide one of the largest and deepest cluster samples with mass calibration available to date. 
\end{section}

\section*{Acknowledgements}
We thank the anonymous referee for the useful comments. FB, MRo and LM thank the support from the grants ASI n.I/023/12/0 ``Attivit\`a relative alla fase B2/C per la missione Euclid'' and PRIN MIUR 2015 ``Cosmology and Fundamental Physics: Illuminating the Dark Universe with Euclid''. MM was supported by the SFB-Transregio TR33 ``The Dark Universe''. MS acknowledges support from the contracts ASI-INAF I/009/10/0, NARO15 ASI-INAF I/037/12/0, ASI 2015-046-R.0 and ASI-INAF n.2017-14-H.0.  NRN and FG acknowledge financial support from the European Unions Horizon 2020 research and innovation programme under the Marie Skodowska-Curie grant agreement No 721463 to the SUNDIAL ITN network.

\bibliographystyle{mnras}
\bibliography{cluster_wl}

\appendix

\section{Background selection}\label{sect:app_backsel}

\begin{figure}
 \includegraphics[width=\columnwidth]{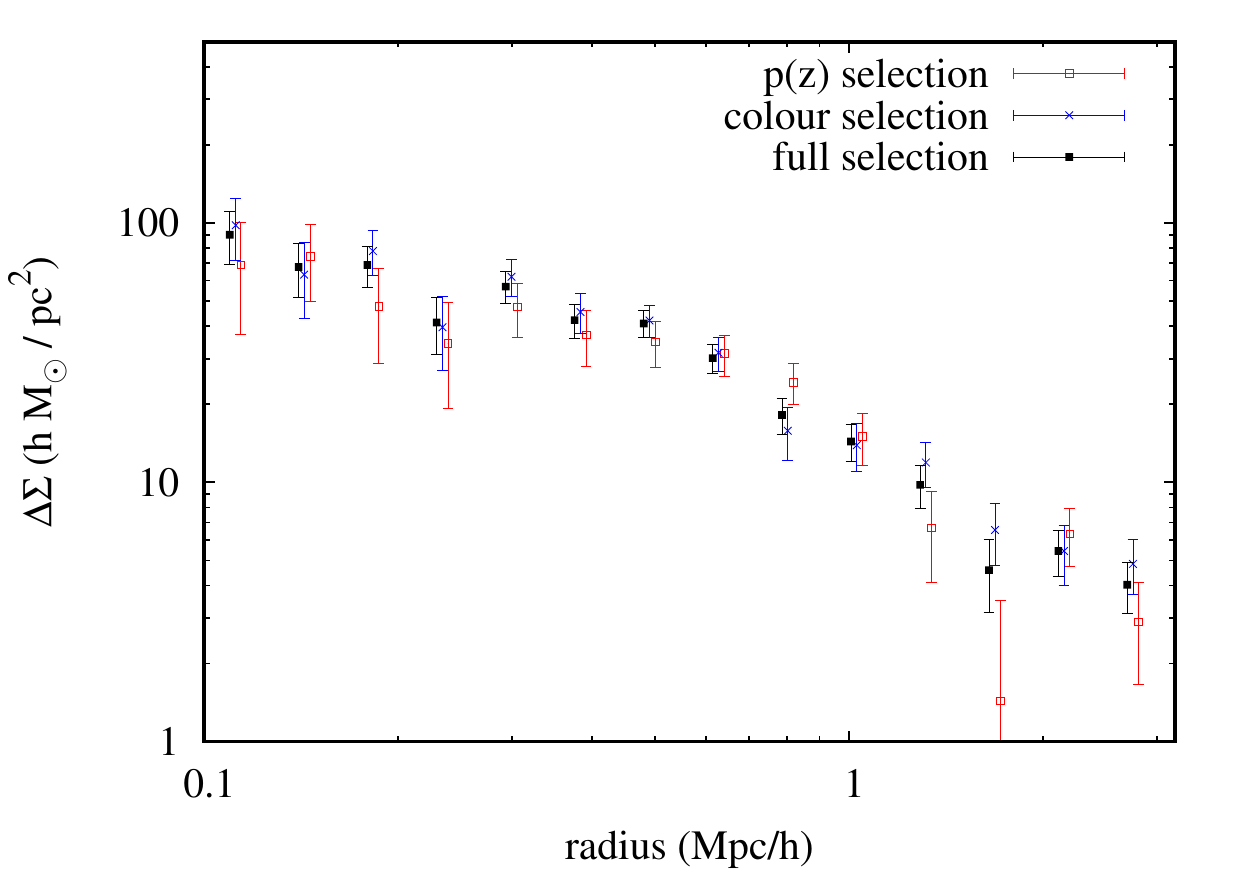}
 \caption{Total $\Delta \Sigma(R)$ for all the clusters at 0.1 < $z$ < 0.3 in the catalogue, for the different background selection criteria: p(z) selection in red, colour selection in blue. In black, the signal from the full background sample. The data points are slightly offset along the x-axis for clarity.}
 \label{fig:back_sel}
\end{figure}

The selection of background galaxies described in Section \ref{sect:sel_back} is done with two complementary methods: we select galaxies with robust photometric redshifts in a favourable $z$ range, and those lying in a high-$z$ region of the $gri$ colour-colour plot. As an additional check, we compare the excess density obtained by employing only one method at a time, as shown in Figure \ref{fig:back_sel}. The figure refers to the stacked profile for all clusters with 0.1 < $z$ < 0.3. The shear profiles are clearly consistent with each other at all radii. For higher redshift clusters, the signal obtained by selecting galaxies through the p(z) is almost negligible in comparison to the colour-colour selection, and this makes the comparison less meaningful.

\section{Lens-source angular cross-correlation}\label{sect:app_cc}

\begin{figure}
 \includegraphics[width=\columnwidth]{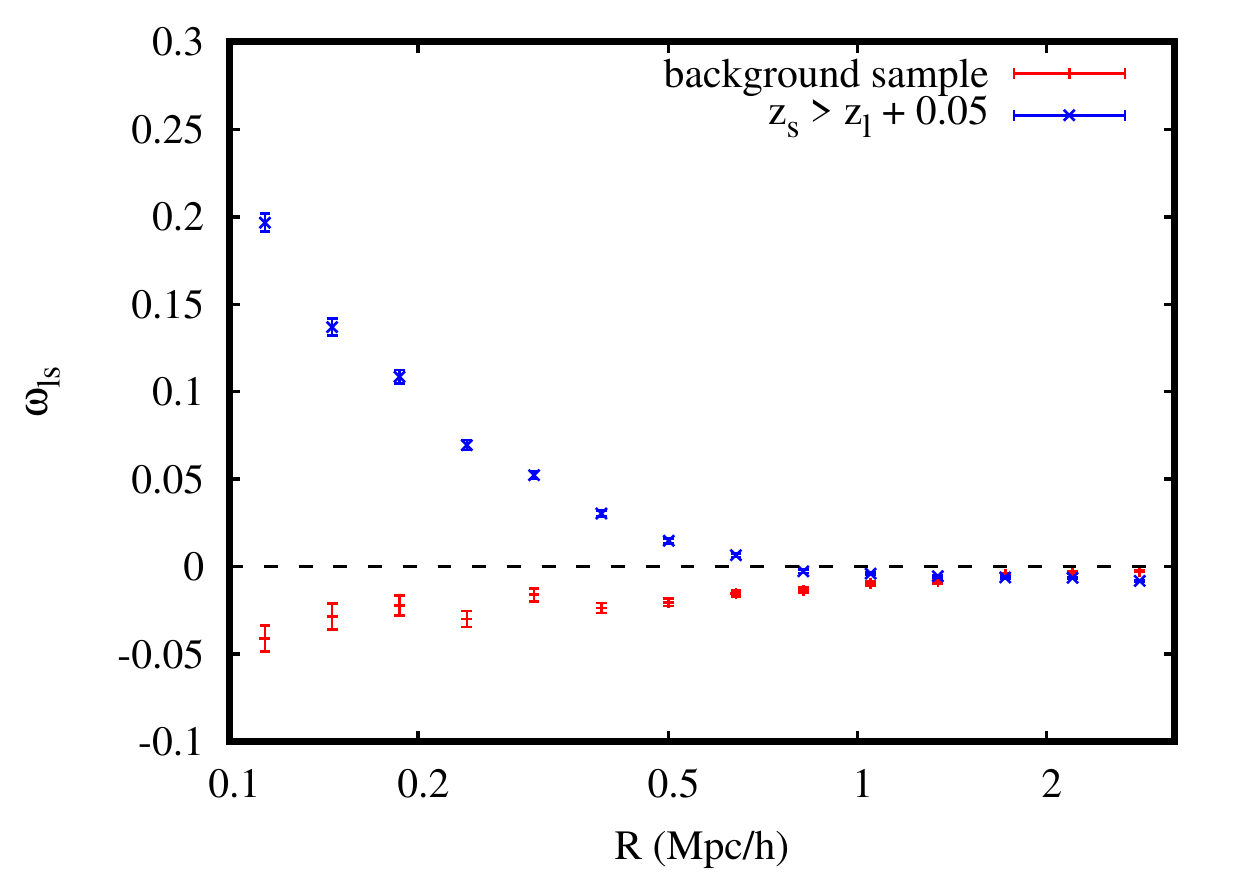}
 \caption{Radial cross correlation $\omega_{\text{ls}}$ between lenses and sources for the background sample we used in the analysis (red points) and for a source selection based on the most likely redshift only (in blue). The errors are estimated assuming a Poissonian statistics on the counts. The dashed horizontal line marks $\omega_{\text{ls}}$ = 0.}
 \label{fig:app_cc}
\end{figure}

In principle, the spatial distribution of background sources should be uncorrelated from the position of the clusters, if physically associated galaxies are excluded. If some cluster galaxies enter the background sample, the number density of background sources should increase towards the lens centre \citep{2014MNRAS.439...48A,2015MNRAS.449.2219M}. Due to our conservative selection of background galaxies, we do not expect a significant contamination from cluster galaxies, as confirmed by the tests performed in Section \ref{sect:syst_back_pz}. As an additional test, we compute the radial cross-correlation $\omega_{\text{ls}} (R)$ between lenses and background sources, making use of a random shear catalogue based on the KiDS-450 masks. The radial cross-correlation is estimated as
\begin{equation}\label{eq:cc}
\omega_{\text{ls}} (R) = \frac {N_{\text{rand}}}{N_{\text{src}}} \frac {C_{\text{ls}} (R)} {C_{\text{lr}} (R)} - 1,
\end{equation}
where $N_{\text{src}}$ and $N_{\text{rand}}$ are the total number of galaxies in the background and in the random sample, respectively, and $C_{\text{ls}} (R)$ and $C_{\text{lr}} (R)$ are the number of galaxies at a radial projected distance $R$ from a lens in the background and in the random sample, respectively. 

In Figure \ref{fig:app_cc} we show in red the results for the background sample we used in the analysis. The radial distribution of background sources is compatible with being uncorrelated with the lens positions at percent level. The observable decrement towards the center of the cluster is likely to be due to the obscuration of background sources by cluster galaxies \citep{2014MNRAS.439...48A}. For comparison, in blue we show the cross-correlation for a different source selection: instead of the background sample, we calculated Equation \ref{eq:cc} for all the sources with redshift $z_\text{s} > z_\text{l} + 0.05$. In this case, we obtain the expected increment towards the center as more cluster galaxies with an over-estimated redshift $z_\text{s}$ enter in $C_{\text{ls}}$. This test confirms that our selection of the background sample removes a significant contamination from cluster galaxies.

\section{Null tests}\label{sect:app_null}

\begin{figure}
 \includegraphics[width=\columnwidth]{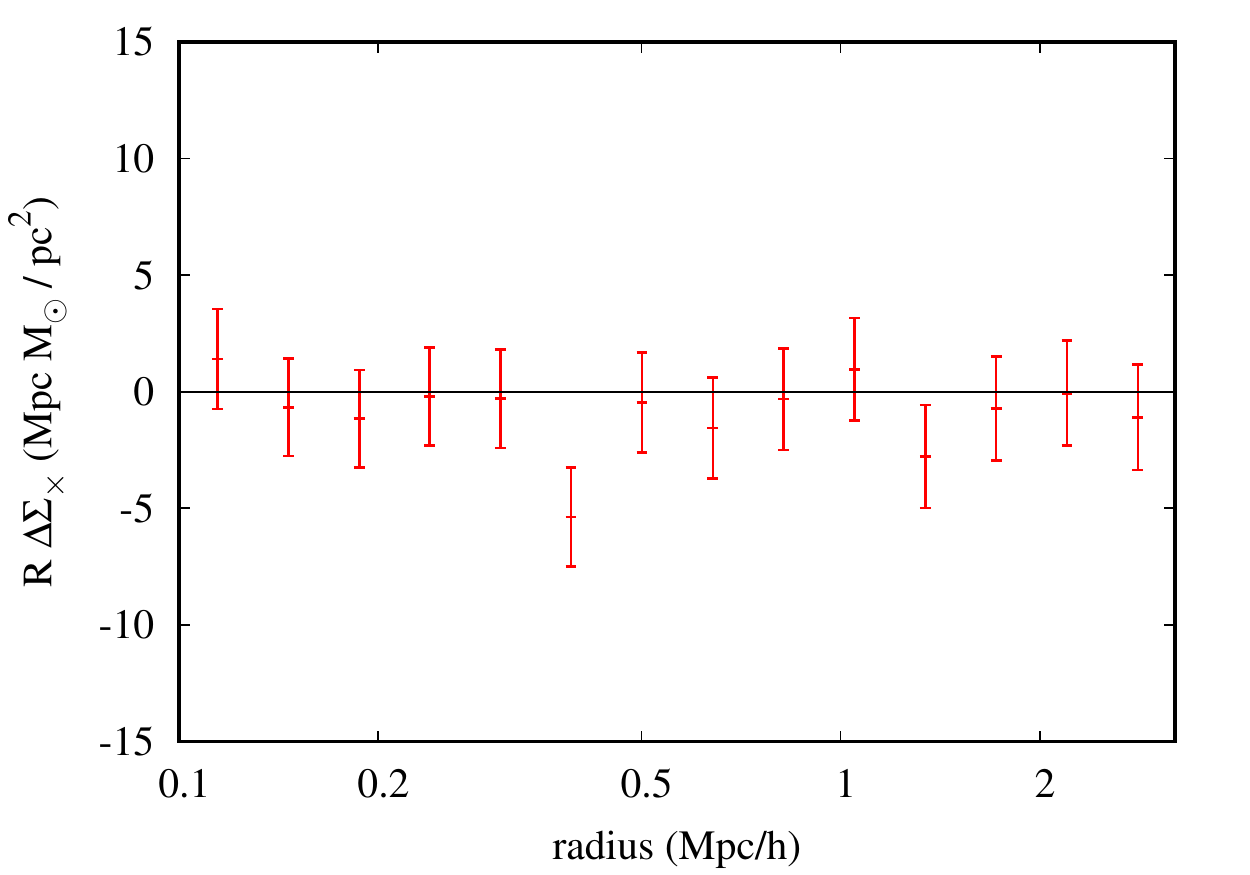}
 \caption{Total $R \Delta \Sigma_\times (R)$, obtained by stacking all the clusters in the catalogue. Vertical error bars represent the square root of the diagonal elements of the covariance matrix.}
 \label{fig:stack_cross}
\end{figure}

To exclude any significant contamination of the recovered profiles by non-lensing signal, which would arise e.g. from an imperfect modelling of the PSF, we compute the cross-component of the stacked shear signal, defined as $\Delta \Sigma_\times (R) = \Sigma_{\text{crit}} \gamma_\times$, where $\gamma_\times$ is the component of the shear directed at 45 $\deg$ from the center of the cluster. This so-called B-mode should be consistent with zero if the signal we measure comes from gravitational lensing. The results are shown in Figure \ref{fig:stack_cross}, where we multiplied $\Delta \Sigma_\times$ by $R$, for visualisation purposes. The signal does not show any significant deviation from zero over all the radial range considered in the analysis. Assuming a null signal, the resulting $\chi^2$ value is 9.94 for 14 degrees of freedom.

\begin{figure}
 \includegraphics[width=\columnwidth]{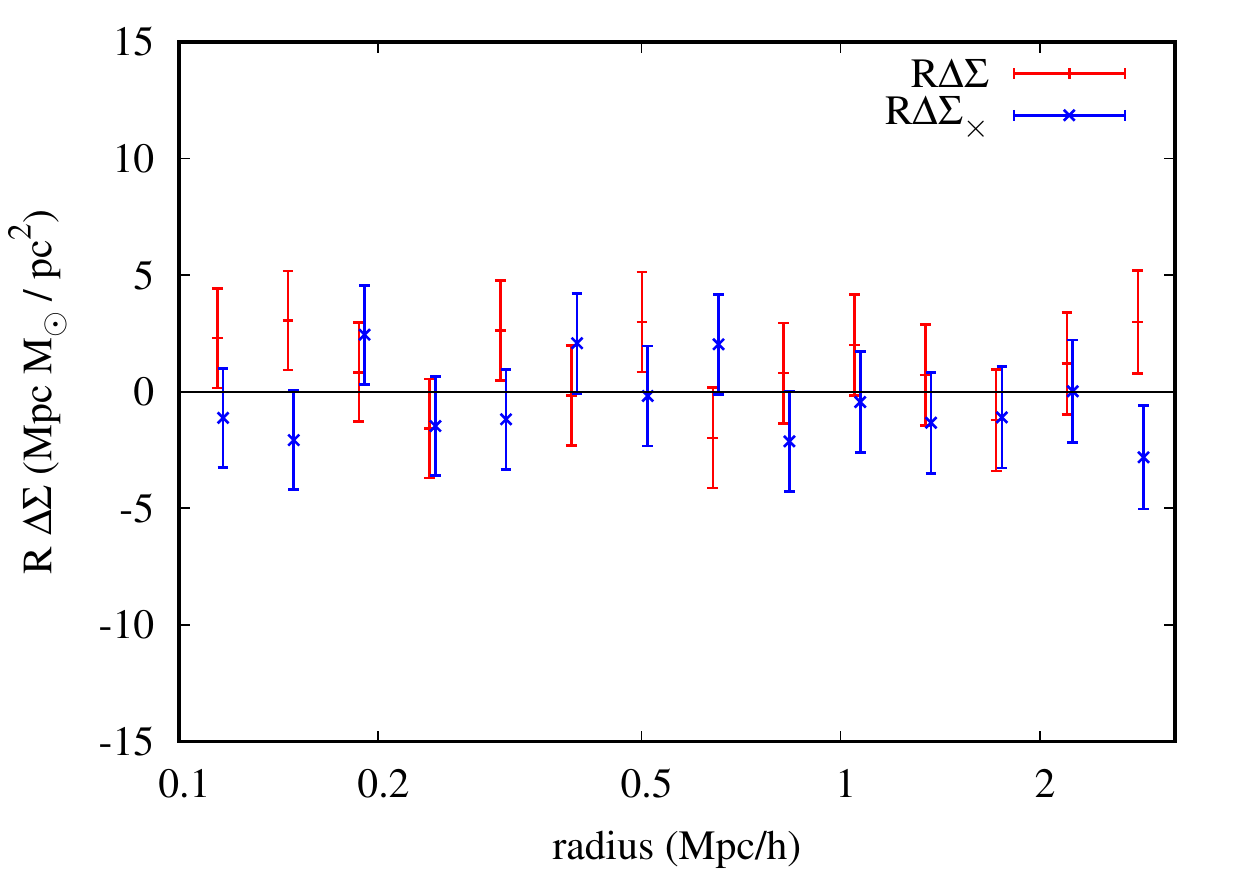}
 \caption{In red, the stacked $R \Delta \Sigma (R)$ around random points with the same redshift distribution as the cluster sample. In blue, the corresponding stacked $R \Delta \Sigma_\times (R)$. Points have been slightly shifted on the x-axis for visibility. Vertical error bars represent the square root of the diagonal elements of the covariance matrix.}
 \label{fig:stack_random}
\end{figure}

Another way to check for systematics is to measure the shear profile around random points in the data field. To this purpose, we created a mock cluster catalogue with the same redshift and amplitude distribution as the one presented in Section \ref{sect:clu_cat}, but with random positions inside the survey footprint. We then repeated the analysis performed on real objects and we extracted the stacked shear profile and its cross-component, shown in Figure \ref{fig:stack_random}. Also in this case, the signal is compatible with zero, indicating there is no residual systematics in the data. Assuming a null signal, the resulting $\chi^2$ values are 11.77 and 8.44 for 14 degrees of freedom. The cross-component around random position is in remarkable agreement with the measurement at the cluster position, further stressing that systematics are well under control.

\bsp	
\label{lastpage}
\end{document}